\documentclass[showkeys,showpacs,prb,eqsecnum]{revtex4}

\usepackage{amsmath,amssymb}
\usepackage[dvips]{graphicx,color}
\usepackage{bm}

\begin{document}

\title{
Anisotropic ground states of the quantum Hall system with currents
}

\author{Kazumi Tsuda}
\author{Nobuki Maeda}
\author{Kenzo Ishikawa}
\affiliation{Department of Physics, Hokkaido University, Sapporo
060-0810, Japan}
\date{\today}

\begin{abstract}
Anisotropic states at half-filled higher Landau levels are 
investigated in the system with a finite electric current. 
We study the response of the striped Hall state and the anisotropic 
charge density wave (ACDW) state against the injected current 
using the effective action.
Current distributions and a current dependence of the 
total energy are determined for both states. 
With no injected current, 
the energy of the ACDW state is lower 
than that of the striped Hall state.
We find that the energy of the ACDW state increases faster than 
that of the striped Hall state as the injected current increases. 
Hence, the striped Hall state becomes the lower energy state 
when the current exceeds the critical value. 
The critical value is estimated at about $0.04$ - $0.07$ nA, 
which is much smaller than the current used in the experiments.
\end{abstract}

\maketitle
\section{Introduction}
In the two-dimensional (2D) electron system  subjected to a
strong perpendicular magnetic field, 
which is called the quantum Hall system,  
the half-filled states at each Landau
level (LL) exhibit much attractive features. 
Around the half-filled lowest LL, 
isotropic compressible states, 
which are widely believed to be the Fermi liquid of composite fermions, 
have been observed.\cite{Composite0,Composite1} 
Around the half-filled second LL, 
the $5/2$ fractionally quantized Hall conductance has been observed.
\cite{5/2} 
The p-wave Cooper pairing state of composite fermions, 
which is called the Pfaffian state, 
has been proposed to explain this state.\cite{Pfaffian1,Pfaffian2} 
Around the half-filled third and higher LLs, 
highly anisotropic states, which have extremely anisotropic longitudinal 
resistivities 
and un-quantized Hall resistivities, 
have been found in ultra-high mobility samples at low
temperature.\cite{stripe_ex1,stripe_ex2} 
Many theoretical works have been done to study the anisotropic states.
\cite{stripe0,stripe00,stripe01,stripe02,stripe03,stripe04,stripe05,
stripe06,stripe07,stripe08,stripe09,ACDW1} 
In the present paper, we focus on 
two different Hartree-Fock (HF) states, i.e., 
a unidirectional charge density wave state,\cite{stripe0,stripe00} 
which is called a striped Hall state in the present paper, 
and an anisotropic charge density wave (ACDW) state.\cite{ACDW1} 

The experimental features of the anisotropic states  
suggest that 
the anisotropic states are 
the striped Hall states. 
The striped Hall state has the anisotropic Fermi surface, which 
has an energy gap in one direction and is gapless in the other direction. 
The anisotropic longitudinal 
resistivities and the un-quantized Hall resistivities 
are naturally explained by this 
anisotropic Fermi surface.\cite{stripe1,stripe2} 
On the other hand, 
the ACDW state has energy gaps in both directions so that 
it is difficult to explain the experiments with the ACDW state.  
However, 
the ACDW state has a lower energy than the striped Hall state 
in the system with no electric current.  
This has been a contradiction between the experiments and the
theories for the anisotropic state. 

In the experiments of the anisotropic states, 
current is injected. 
This current effect has not been taken into account 
in the previous calculations of the total energy. 
MacDonald et al. have studied 
the injected current effect on the integer quantum Hall system
about two decades ago.\cite{Mac1} 
They have calculated the current and charge distributions 
and found 
that charges accumulate around both edges 
of the sample with the opposite sign, 
as expected from the classical Hall effect.\cite{Mac1,Mac2,Mac3,Mac4} 
The charge accumulation causes the energy enhancement via the 
Coulomb interaction between charged particles. 
The same type of energy corrections may exist even in 
highly correlated quantum Hall states. 
However, the effect of the injected current on the anisotropic state 
has not been studied.

In the present paper, we calculate the correlation energies of 
the striped Hall state and the ACDW state in the system with the injected current, 
no impurities, and no metallic contacts. 
It is important to know if the ACDW state has a lower energy even in the
system with the injected current. 
For this purpose, the dependence of the correlation energies on the
injected current is studied in detail. 
Effects of impurities and metallic contacts are ignored in our
calculations of the correlation energies since these effects are 
expected to be small in the experiments of the anisotropic states in the 
ultra-high mobility samples and are outside the scope of this work. 
The effects of the injected current are investigated 
using the response functions for electromagnetic fields. 
The current and charge distributions are determined and  
the energies of the two states are calculated from these distributions. 
It is found that the energy of the ACDW state  
increases faster than that of the striped Hall state 
as the injected current increases. 
Hence, the striped Hall state becomes the lower energy state 
when the current exceeds the critical value. 
The critical value is estimated at about $0.04$ - $0.07$ nA, 
which is much smaller than the current used in the experiments. 
Our result suggests that the anisotropic states observed in the
experiments are the striped Hall states. 
Hence, the contradiction between the experiments and the theories is
resolved.

This paper is organized as follows. 
In Sec. \ref{sec:HF}, the two HF states, i.e., 
the striped Hall state and the ACDW state, 
are constructed in the von Neumann lattice formalism. 
In Sec. \ref{sec:response}, electromagnetic response functions 
of the two HF states are calculated in the long wavelength limit. 
Using these response functions, 
we determine the current and charge distributions and calculate 
the energy corrections due to currents 
in Sec. \ref{sec:energy}. 
A summary is given in Sec. \ref{sec:summary}.

\section{Hartree-Fock ground states on the von Neumann lattice}
\label{sec:HF}
In this section, the striped Hall state and the ACDW state are
constructed in the HF approximation using the von Neumann lattice base. 
The von Neumann lattice base is suitable for studying spatially periodic
states. 
We first review the von Neumann lattice base for 
the completeness of the present paper. 

\subsection{von Neumann lattice base}
Let us consider the 2D electron system 
in a uniform external magnetic field 
$B=(\nabla \times {\bf A})_z$.
The spin degree of freedom is ignored and the natural unit
$(\hbar=c=1)$ is used in the present paper. 
We introduce two sets of coordinates, i.e., 
the relative coordinates ${\bm \xi}=(\xi,\eta)$ 
and the guiding center coordinates ${\bf X}=(X,Y)$:
\begin{gather}
 \xi=\frac{1}{eB}(-i\partial_y+eA_y),\qquad 
 \eta=-\frac{1}{eB}(-i\partial_x+eA_x), \notag\\
 X=x-\xi,\qquad Y=y-\eta, 
\end{gather}
where $e>0$.
Each set of coordinates satisfies the canonical commutation 
relations:
\begin{gather}
 [X,Y]=-[\xi,\eta]=i/eB \notag\\
 [X,\xi]=[X,\eta]=[Y,\xi]=[Y,\eta]=0.
\end{gather}
Using these variables, 
the one-particle free Hamiltonian is written in the form 
\begin{equation}
 H_0=\frac{1}{2}m\omega_c^2(\xi^2+\eta^2), 
\end{equation}
where $\omega_c=eB/m$ is a cyclotron frequency. 
Since $H_0$ is equivalent to the Hamiltonian of a harmonic oscillator, 
the eigenvalue splits into each LL as follows: 
\begin{equation}
 H_0|f_l\rangle = E_l| f_l\rangle,\quad
  E_l=\omega_c(l+\frac{1}{2}),\quad (l=0,1,2\dots).
\end{equation}
 
It is convenient to use a discrete set of coherent states 
of guiding center coordinates, 
\begin{equation}
 (X+iY)|\alpha_{mn}\rangle =z_{mn}|\alpha_{mn}\rangle,\quad 
 z_{mn}=a(r_sm +i\frac{n}{r_s}), 
\end{equation}
where $m$ and $n$ are integers. 
The completeness of the set $\{|\alpha_{mn}\rangle \}$ is 
ensured \cite{vNL0,vNL1},  
and this set is called the von Neumann lattice (vNL) base.\cite{vNL2} 
These coherent states are localized at the rectangular lattice point 
$a(mr_s, n/r_s)$, 
where a positive real number $r_s$ is an asymmetry parameter of the unit cell 
and $a=\sqrt{2\pi/eB}$ is a lattice constant. 
We set $a=1$ unless otherwise stated. 
Note that the number of lattice points is equal to the number of
states in one LL. 
By Fourier transforming these states, we obtain the orthonormal basis in
the momentum representation, 
\begin{gather}
 |\beta_{\bf p}\rangle =\sum_{m,n}e^{ip_x m+ip_y n}|\alpha_{mn}\rangle
 /\beta({\bf p}),\notag \\
 \beta({\bf p})=(\sqrt{2}\,r_s)^{1/2}e^{-(r_sp_y)^2/4\pi}
 \vartheta_1\left( \frac{p_x+ir_s^2p_y}{2\pi}\Biggm|ir_s^2 \right),
 \notag \\
 \langle \beta_{\bf p}| \beta_{{\bf p}'} \rangle =
 \sum_{\bf N}(2\pi)^2\delta^2({\bf p}-{\bf p}'-2\pi{\bf N})e^{i\phi({\bf
 p}',{\bf N})}, 
\end{gather}
where $\vartheta_1$ is a Jacobi's theta function of the first kind, 
${\bf N}=(N_x,N_y)$ is a vector with integer values, and 
$\phi({\bf p},{\bf N})=\pi(N_x+N_y)-p_xN_y$.  
The two-dimensional lattice momentum ${\bf p}$ is defined 
in the Brillouin zone (BZ), $|p_i|<\pi$, and 
$\beta({\bf p})$ obeys a nontrivial boundary condition
\begin{equation}
 \beta({\bf p}+2\pi{\bf N})=e^{i\phi({\bf p},{\bf N})}\beta({\bf p}). 
  \label{eq:BC_for_beta}
\end{equation} 
The Hilbert space of a one-particle state is spanned by the state
$|l,{\bf p}\rangle=|f_l\rangle \otimes |\beta_{\bf p}\rangle$. 
We use this base throughout our calculation. 

The electron field operator is expanded by the vNL base as
\begin{equation}
\Psi({\bf x})=\sum^\infty_{l=0}
 \int_{{\rm BZ}}\frac{d^2p}{(2\pi)^2}b_l({\bf p})\langle {\bf
 x}|l,{\bf p}\rangle, 
 \label{eq:vNL}
\end{equation}
where 
$b_l({\bf p})$ obeys the same boundary condition as
Eq.(\ref{eq:BC_for_beta}) 
and satisfies the following anti-commutation relation: 
\begin{equation}
 \{b_l({\bf p}), b^\dag_{l'}({\bf p}')\}=\delta_{l,l'}\sum_{\bf N}(2\pi)^2
  \delta^2({\bf p}-{\bf p}'-2\pi{\bf N})e^{i\phi({\bf p},{\bf N})}. 
\end{equation} 
The Fourier transform of the density operator 
$\rho({\bf k})
=\int d^2x e^{i{\bf k}\cdot {\bf x}}\Psi^\dag({\bf x})\Psi({\bf x})$ 
is written as 
\begin{align}
 \label{eq:expression_of_rho}
 \rho({\bf k})=&\sum_{l,l'}
  \int_{\rm BZ}\frac{d^2p}{(2\pi)^2}b_l^\dag({\bf p})
  b_{l'}({\bf p}-\hat{\bf k})
  f^0_{l,l'}({\bf k}) e^{-(i/4\pi)\hat{k}_x(2p_y-\hat{k}_y)},
\end{align}
where 
$f_{l,l'}^0({\bf k})=\langle f_l| e^{i{\bf k}\cdot{\bm \xi}}|f_{l'}\rangle$
(see Appendix \ref{app:LL_matrix}) and 
$\hat{\bf k}=(r_sk_x,k_y/r_s)$. 

For a strong magnetic field, 
in which the energy difference 
between the nearest LLs, $\omega_c$, 
is much larger than the typical order of the Coulomb interaction, 
$e^2/4\pi\epsilon l_B$, where $\epsilon$ is the dielectric constant and 
$l_B=\sqrt{1/eB}$ is the magnetic length, 
the LL mixing effects can be neglected. 
In this case, the Hamiltonian is projected to the uppermost
partially-filled LL, and the kinetic term is quenched. 
The Hamiltonian in this system is given by the projected Coulomb
interaction, 
\begin{gather}
 H^{(l)}_{{\rm int}}=\frac{1}{2} \int \frac{d^2 k}{(2\pi )^2} 
  : \rho_l({\bf k}) V({\bf k}) \rho_l(-{\bf k}):\notag,\\
 V({\bf k})=\frac{2\pi q^2}{k}\ (k\neq 0, q^2=\frac{e^2}{4\pi\epsilon}),
 \quad V(0)=0. 
 \label{eq:Coulomb}
\end{gather}
In Eq. (\ref{eq:Coulomb}), 
the colons represent a normal ordering with respect to 
creation and annihilation operators, 
$l$ denotes the uppermost partially-filled LL index, 
and $\rho_l({\bf k})$ 
is given by 
\begin{gather}
 \rho_l({\bf k})=f^0_{l,l}({\bf k})\bar{\rho}_l({\bf k})\notag \\
 f_{l,l}^0({\bf k})=e^{-k^2/8\pi} 
 L^0_l(\frac{k^2}{4\pi}), \notag \\
 \bar{\rho}_l({\bf k})= \int_{{\rm BZ}}\frac{d^2 p}{(2\pi)^2} 
 b^\dag_l({\bf p}) b_l({\bf p}-\hat{{\bf k}})
 e^{-(i/4\pi)\hat{k}_x(2p_y-\hat{k}_y)}, 
\end{gather}
where 
$\bar{\rho}_l({\bf k})$ is a projected density operator and 
$L^0_l$ is a Laguerre polynomial. 

$H_{\rm int}$ 
is expressed 
in the HF approximation as 
(Appendix \ref{app:HF_Hamiltonian}) \cite{HF0, HF}  
\begin{equation}
 H_{\rm HF}^{(l)}=\mathcal{H}_{\rm HF}^{(l)}-\frac{1}{2}
  \langle \mathcal{H}_{\rm HF}^{(l)}\rangle , 
\end{equation}
where 
\begin{equation}
 \label{eq:LL_projected_Hamiltonian}
 \mathcal{H}^{(l)}_{{\rm HF}}=\int\frac{d^2 k}{(2\pi)^2}
 v^{\rm HF}_l(\tilde{\bf k})\langle
 \bar{\rho}_l(-\tilde{\bf k})\rangle\bar{\rho}_l(\tilde{\bf k}),
\end{equation}
\begin{gather}
 v^{\rm HF}_l({\bf k})=v_l({\bf k})-\int\frac{d^2k'}{(2\pi)^2}
 v_l({\bf k}')e^{(i/2\pi)(k'_xk_y-k'_yk_x)}, \notag\\
 v_l({\bf k})=V({\bf k})(f_{l,l}^0({\bf k}))^2, \quad 
 \tilde{{\bf k}}=(\frac{k_x}{r_s}, r_s k_y).  
\end{gather}
This HF Hamiltonian has been 
diagonalized self-consistently, and 
various ground states have been obtained. 
In the present paper, 
we concentrate on the striped Hall state and 
the ACDW state at half-filled higher LLs. 
These states are constructed using the vNL base in the following
subsections. 

\subsection{Striped Hall state}
We consider the case of the filling factor $\nu=l+\nu^\ast$, with
$\nu^\ast=1/2$. The present formalism is valid  for the arbitrary
$\nu^\ast$ ($0< \nu^\ast < 1$). 
The striped Hall state is a unidirectional charge density wave state 
which has the following unidirectional density
(Fig. \ref{fig:dens_stripeL2}): 
\begin{figure}
\includegraphics[width=6cm]{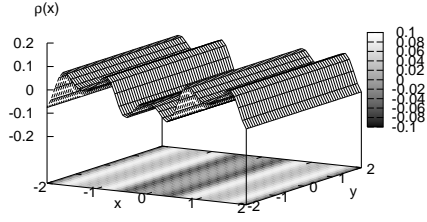}
\caption{\label{fig:dens_stripeL2}Density of the 
$l=2$ striped Hall state at half-filling. 
The uniform part $\rho_0=\nu^\ast$ is subtracted. 
The density of the striped Hall state 
is uniform in the $y$-direction (stripe direction) 
and periodic with a period $r_s$ 
in the $x$-direction (perpendicular direction).}
\end{figure}
\begin{equation}
 \langle \rho_l({\bf x}) \rangle_{\rm
  stripe}=\sum_{N_x}\Delta_l(N_x)f_{l,l}^{0}(\frac{2\pi N_x}{r_0}, 0)
  e^{i(2\pi N_x /r_0)x},
  \label{eq:stripe_dens}
\end{equation}
where $r_0$ is the period of the density in the $x$-direction, 
$\Delta_l(N_x)$ is an order parameter
determined self-consistently, and $\Delta_l(0)=\nu^\ast$. 
We call the uniform direction {\it stripe direction} and 
the other direction {\it perpendicular direction} in this paper. 
Eqation (\ref{eq:stripe_dens}) gives the following form of 
$\langle \bar{\rho}_l({\bf k}) \rangle$:
\begin{equation}
 \label{eq:assumption_stripe}
 \langle \bar{\rho}_l({\bf k}) \rangle _{\rm stripe}=
  \sum_{N_x}\Delta_l(N_x)(2\pi)^2
  \delta(k_x+\frac{2\pi N_x}{r_0})\delta(k_y).  
\end{equation}
The HF Hamiltonian of the striped Hall state is easily diagonalized on the vNL 
by taking the asymmetry parameter $r_s=r_0$. 
Substituting Eq. (\ref{eq:assumption_stripe}) into the HF Hamiltonian
and using $r_s=r_0$, the HF Hamiltonian of the striped Hall state is
written by  
\begin{equation}
 \mathcal{H}_{\rm HF-stripe}^{(l)}=\int _{\rm BZ}\frac{d^2 p}{(2\pi)^2}
  \epsilon_l({\bf p})b^\dag_l({\bf p})b_l({\bf p}), 
  \label{eq:Hamiltonian_stripe}
\end{equation}
where $\epsilon_l({\bf p})$ is a one-particle energy given by 
\begin{equation}
 \epsilon_l({\bf p})=\epsilon_0+\sum_{N_x\neq 0}\Delta_l(N_x)
  v_l^{\rm HF}(\frac{2\pi N_x}{r_s}, 0)(-1)^{N_x}e^{-iN_xp_y}. 
  \label{eq:one-particle_ene_stripe}
\end{equation} 
In Eq. (\ref{eq:one-particle_ene_stripe}), $\epsilon_0$ is a uniform
Fock energy given by $\nu^\ast v^{\rm HF}_l(0)$.

\begin{figure}
 \includegraphics[width=4cm]{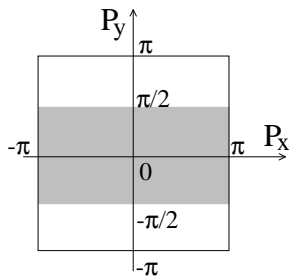}
 \caption{\label{fig:FS}
Fermi sea of the striped Hall state at half-filling. 
The occupied state is represented by the dark region. 
When the stripe direction faces the $y$-direction, the $p_x$-direction 
of the Brillouin zone is fully occupied. 
In this case, the Fermi sea has the inter-LL energy gap in the
 $p_x$-direction and is gapless in the $p_y$-direction.} 
\end{figure}

\begin{figure}
 \includegraphics[width=6cm]{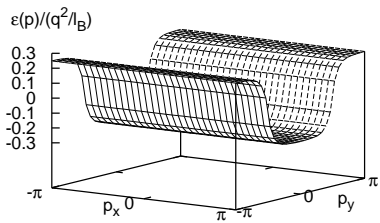}
 \caption{\label{fig:ene_stripeL2}
One-particle energy of the $l=2$ 
striped Hall state at 
half-filling. 
The uniform Fock energy is subtracted. 
When the stripe direction faces the $y$-direction, 
the one-particle energy is uniform in the $p_x$-direction.}
\end{figure}

From Eq. (\ref{eq:Hamiltonian_stripe}), the two-point function of the
operator $b_l({\bf p})$ is given by 
\cite{stripe1,stripe2} 
\begin{align}
 \langle b^\dag_l({\bf p})b_{l'}({\bf p}')\rangle_{\rm stripe}
 =&\sum_{\bf N}\delta_{l,l'}\theta[\mu_{\rm F}-\epsilon_l({\bf p})]
 (2\pi)^2\delta^2({\bf p}-{\bf p}'-2\pi{\bf N})
 e^{-i\phi({\bf p},{\bf N})}, 
 \label{eq:two_point_stripe}
\end{align}
where $\mu_{\rm F}$ is a Fermi energy and $\theta$ is a step function. 
The self-consistent equation for $\Delta_l(N_x)$ is obtained by 
substituting Eq. (\ref{eq:two_point_stripe}) into the left hand side of
Eq. (\ref{eq:assumption_stripe}).
$\Delta_l(N_x)=(-1)^{N_x}\sin(\nu^\ast\pi N_x)/\pi N_x$ 
is a solution of the self-consistent equation. 
This solution has the Fermi sea, $|p_y|<\pi\nu^\ast$  
(shown in Fig. \ref{fig:FS}) 
and gives the one-particle energy as 
(Fig. \ref{fig:ene_stripeL2})
%
\begin{equation}
 \epsilon_l({\bf p})=\epsilon_0+\sum_{N_x\neq 0}v_l^{\rm HF}(\frac{2\pi N_x}{r_s},
  0)\frac{\sin(\nu^\ast\pi N_x)}{\pi N_x}e^{-iN_x p_y}. 
\end{equation}
The HF energy per particle is given as a function of $r_s$ by 
\begin{align}
 E_{\rm stripe}^{(l)}(r_s)=
  \frac{\langle H_{\rm HF}^{(l)}\rangle_{\rm
 stripe}}{N_e^{(l)}}
  =\frac{1}{2}\epsilon_0
  +\frac{1}{2}\sum_{N_x\neq 0}\nu^\ast 
  v_l^{\rm HF}(\frac{2\pi N_x}{r_s}, 0) 
  \left(\frac{\sin(\nu^\ast\pi N_x)}{\nu^\ast\pi N_x}\right)^2. 
\end{align}
where $N_e^{(l)}$ is the total number of particles within the $l$th LL. 
The optimal value of $r_s$ is determined by minimizing $E_{\rm stripe}^{(l)}(r_s)$. 
The optimal value of $r_s$ and the minimum energy at each LL are
shown in Table \ref{table:stripe_table}.\cite{stripe2} 
\begin{table}
\caption{Minimum energy and corresponding parameter $r_s$ of the 
striped Hall states at $\nu=l+1/2$.}
\label{table:stripe_table}
\begin{ruledtabular}
\begin{tabular}{ccc}
$l$ & $r_s^{\rm stripe}$ & $E_{\rm stripe}/(q^2/l_{\rm B})$\\
\colrule
0 & 1.636 & -0.4331 \\
1 & 2.021 & -0.3490 \\
2 & 2.474 & -0.3074 \\
3 & 2.875 & -0.2800 \\
\end{tabular}
\end{ruledtabular}
\end{table}

The striped Hall state has the anisotropic Fermi surface shown in Fig. \ref{fig:FS}, 
which has an inter-LL energy gap in the $p_x$-direction and is gapless
in the $p_y$-direction. This would cause the anisotropic longitudinal
resistivities. 

\subsection{Anisotropic charge density wave state}\label{sec:ACDW}
\begin{figure}
\includegraphics[width=6cm]{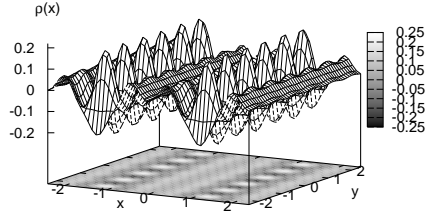}
\caption{\label{fig:dens_ACDWL2}Density of the $l=2$ ACDW state at 
half-filling. 
The uniform part $\rho_0=\nu^\ast$ is subtracted. 
The density of the ACDW state 
is periodic in both directions. The $y$-direction 
is referred to as {\it ACDW direction} in this figure.} 
\end{figure}

We consider the ACDW state with the following rectangular charge
density wave (Fig. \ref{fig:dens_ACDWL2}) : \cite{ACDW0}
%
\begin{figure}
\includegraphics[width=4cm]{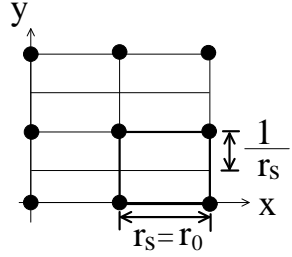}
\caption{\label{fig:Lattice}
vNL unit cell and ACDW unit cell. 
The black circles represent the location of electrons 
and the thin lines represent the vNL. 
The ACDW unit cell represented by the bold lines 
is just twice as large as the vNL unit cell 
in the case of $\nu^\ast=1/2$.}
\end{figure}
\begin{equation}
 \langle \rho_l({\bf x}) \rangle_{\rm ACDW}=
  \sum_{\bf N}\Delta_l({\bf Q}_N)f_{l,l}^{(l)}({\bf Q}_N)e^{-i{\bf Q}_N\cdot
  {\bf x}}, 
  \label{eq:density_ACDW}
\end{equation}
where ${\bf Q}_N=(2\pi N_x/r_{0x}, 2\pi N_y/r_{0y})$, 
$r_{0x}$ and $r_{0y}$ are the periods of the density in the $x$-direction
and the $y$-direction respectively, 
$\Delta_l({\bf Q}_N)$ is an order parameter determined
self-consistently, 
and $\Delta_l(0)=\nu^\ast$. 
We call the direction with a shorter period of the density {\it ACDW
direction} in this paper. 
Equation (\ref{eq:density_ACDW}) gives the following form of 
$\langle \bar{\rho}_l({\bf k})\rangle$: 
\begin{equation}
 \langle \bar{\rho}_l({\bf k}) \rangle_{\rm ACDW}=
  \sum_{\bf N}
  \Delta_l({\bf Q}_N)(2\pi)^2\delta^2({\bf k}-{\bf Q}_N). 
  \label{eq:CDW}
\end{equation}
The number of ACDW unit cells $N_{\rm CDW}=(area)/r_{0x}r_{0y}$
is equal to the number of electrons within the $l$th LL, 
while the number of vNL unit cells $N_{\rm vNL}=(area)/a^2$ 
is equal to the number of states in one
LL.
Here, we write the vNL constant $a$ explicitly. 
Hence, the filing factor $\nu^\ast$ is expressed as  
\begin{equation}
 \nu^\ast=\frac{N_{\rm CDW}}{N_{\rm vNL}}=\frac{a^2}{r_{0x}r_{0y}}. 
\end{equation}
Using the dimensionless lattice parameter $r_0$
defined as $r_{0x}=ar_0$, 
${\bf Q}_N$ is written as  
\begin{equation}
 {\bf Q}_N=(\frac{2\pi N_x}{ar_0}, \frac{2\pi N_y \nu^\ast r_0}{a}). 
\end{equation} 
and $\langle \bar{\rho}_l({\bf k})\rangle_{\rm ACDW}$ is rewritten as 
\begin{align}
 \langle \bar{\rho}_l({\bf k}) \rangle_{\rm ACDW} =&\sum_{\bf N} 
  \Delta_l(\frac{2\pi N_x}{r_0}, 2\pi N_y \nu^\ast r_0)(2\pi)^2\delta(k_x-\frac{2\pi N_x}{r_0})
  \delta(k_y-2\pi N_y \nu^\ast r_0), 
  \label{eq:CDW2}
\end{align}
where we set $a=1$ again.

\begin{figure}
 \includegraphics[width=6cm]{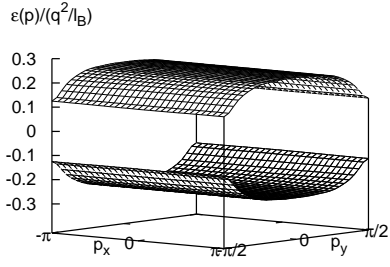}
 \caption{\label{fig:ene_ACDWL2}
One-particle energy of the $l=2$ 
ACDW state at half-filling. 
The uniform Fock energy is subtracted. 
Two bands are formed and  
the lower band is completely filled.}
\end{figure}

The HF Hamiltonian of the ACDW state is easily diagonalized 
using the vNL base. 
We concentrate on the case of $\nu^\ast=1/2$. 
In this case, 
the ACDW unit cell is 
just twice as large as
the vNL unit cell 
(Fig. \ref{fig:Lattice}). 
When we take $r_s=r_0$ and 
divide $N_y$ in the right hand side of Eq. (\ref{eq:CDW2}) into even and odd, 
the mean value of the projected density operator is 
rewritten by 
%
\begin{widetext}
\begin{align}
 \langle \bar{\rho}_l(\tilde{{\bf k}}) \rangle_{\rm ACDW}
 =&\sum_{\bf N}\Big\{ 
 \Delta_l(\frac{2\pi N_x}{r_s}, 2\pi N_yr_s)
 (2\pi)^2\delta^2({\bf k}-2\pi {\bf N}) \nonumber \\
 &+\Delta_l(\frac{2\pi N_x}{r_s},\pi(2N_y+1)r_s)
 (2\pi)^2\delta(k_x-2\pi N_x)\delta(k_y-\pi(2N_y+1))\Big\}. 
 \label{eq:projected_density_of_ACDW}
\end{align}
Substituting this expression into 
Eq. (\ref{eq:LL_projected_Hamiltonian}), 
the HF Hamiltonian of the ACDW state is obtained by 
\begin{equation}
 \mathcal{H}_{\rm HF-ACDW}^{(l)}=\epsilon_0 N_e^{(l)}+
 \int_{\rm BZ}\frac{d^2 p}{(2\pi)^2}
 \left\{
 A({\bf p})b^\dag_l({\bf p})b_l({\bf p})+
 B({\bf p})b^\dag_l({\bf p})b_l(p_x, p_y+\pi)
 \right\}, 
 \label{eq:HF-ACDW0}
\end{equation}
\begin{align}
 A({\bf p})=&\sum_{{\bf N}\neq 0}v_l^{\rm HF}(\frac{2\pi N_x}{r_s}, 2\pi N_yr_s)
 \Delta_l(\frac{2\pi N_x}{r_s}, 2\pi N_yr_s)
 e^{i\pi(N_x+N_y+N_xN_y)-ip_xN_y+ip_yN_x}\nonumber \\
 B({\bf p})=&\sum_{\bf N}v_l^{\rm HF}
 (\frac{2\pi N_x}{r_s},\pi(2N_y+1)r_s)
 \Delta_l(\frac{2\pi N_x}{r_s},\pi(2N_y+1)r_s)
 e^{i\pi(N_x+N_y+N_x(N_y+1/2))-ip_xN_y+ip_yN_x}. 
\end{align} 
\end{widetext}
In this Hamiltonian, 
momentum is not conserved 
since $b_l^\dag({\bf p})$ is coupled with $b_l(p_x, p_y+\pi)$. 
However, 
using the boundary condition for $b_l({\bf p})$, 
the HF Hamiltonian is rewritten as 
\begin{equation}
 \mathcal{H}^{(l)}_{{\rm HF-ACDW}}=\epsilon_0 N_e^{(l)}+
  \int_{{\rm RBZ}}\frac{d^2p}{(2\pi)^2}{\bf
  b}_l^\dag({\bf p})D_l({\bf p}){\bf b}_l({\bf p}),
\label{eq:HF-ACDW}
\end{equation}
\begin{gather}
 {\bf b}_l({\bf p})=
 \left(
  \begin{array}{cc}
   b_l(p_x, p_y)\\
   b_l(p_x, p_y+\pi)
  \end{array}
 \right),\notag \\
 D_l({\bf p})=
 \left(
 \begin{array}{cc}
  A({\bf p}) & B({\bf p})\\
  B^\ast({\bf p}) & A(p_x, p_y+\pi)
 \end{array}
 \right), 
\end{gather}
where the momentum integration is performed over 
the reduced Brillouin zone (RBZ), $|p_x|<\pi$ and $|p_y|<\pi/2$, and 
$D_l({\bf p})$ is a $2\times 2$ Hermite matrix.  
The Hamiltonian expressed by Eq. (\ref{eq:HF-ACDW}) can be diagonalized 
at each momentum just by unitary transforming the field operator 
${\bf b}_l({\bf p})$. 
In the present case of $\nu^\ast=1/2$, 
the Brillouin zone is reduced to the half size of 
the original domain and 
two energy bands are formed (Fig. \ref{fig:ene_ACDWL2}). 
$D_l({\bf p})$ is diagonalized using
the unitary matrix $U({\bf p})$ as
\begin{equation}
 U^\dag({\bf p}) D_l({\bf p})U({\bf p})=
  \left(
   \begin{array}{cc}
    \epsilon_+({\bf p})& 0           \\
    0                & \epsilon_-({\bf p})
   \end{array}
  \right). 
\end{equation}
where $\epsilon_+({\bf p})$ and $\epsilon_-({\bf p})$ represent 
the upper energy band and the lower energy band respectively. 
$\epsilon_{\pm}({\bf p})$ and $U({\bf p})$ are 
given in Appendix \ref{app:self_eq}. 
Using the base ${\bf c}_l({\bf p})=U^\dag({\bf p}){\bf b}_l({\bf p})$, 
the HF Hamiltonian of the ACDW state is obtained by 
\begin{align}
 \mathcal{H}^{(l)}_{{\rm HF-ACDW}}=&\epsilon_0 N_e^{(l)}+
  \int_{{\rm RBZ}}\frac{d^2p}{(2\pi)^2}
  {\bf c}_l^\dag({\bf p})
  \left(
   \begin{array}{cc}
    \epsilon_+({\bf p})& 0           \\
    0                & \epsilon_-({\bf p})
   \end{array}
  \right) 
  {\bf c}_l({\bf p}),
\end{align}
where
\begin{equation}
 {\bf c}_l({\bf p})=
 \left(
  \begin{array}{cc}
   c_+(\bf p)\\
   c_-(\bf p)
  \end{array}
 \right),
\end{equation}
$\Delta_l({\bf Q}_N)$ is determined by solving the self-consistent
equation numerically (see Appendix \ref{app:self_eq}). 

As in the case of the striped Hall state, 
the HF energy of the ACDW state also depends on 
the asymmetry parameter $r_s$. 
The optimal value of
$r_s$, the HF energy per particle, and the magnitude of the energy gap are
given in Table \ref{table:ACDW_table},  
\begin{table}
\caption{\label{table:ACDW_table}
Minimum energy and corresponding parameter $r_s$ 
of the ACDW states at $\nu=l+1/2$. $\Delta_{\rm Gap}$ is a magnitude of the 
energy gap.}
\begin{ruledtabular}
\begin{tabular}{cccc}
$l$ & $r_s^{\rm ACDW}$ & $E_{\rm ACDW}/(q^2/l_{\rm B})$ & 
$\Delta_{\rm Gap}/(q^2/l_{\rm B})$\\
\colrule
0 & $\sqrt{2}$  & -0.4436 &0.3292\\
1 & 1.02, 1.96 & -0.3583 &0.3077\\
2 & 0.82, 2.44 & -0.3097 &0.2470\\
3 & 0.70, 2.86 & -0.2814 &0.1967\\
\end{tabular}
\end{ruledtabular}
\end{table}
in which there are two values of $r_s$ at each
LL due to the $\pi/2$-rotational symmetry. 
The magnitude of the energy gap is estimated to be of the order of $10$ K 
for a few tesla. 
Experiments for the anisotropic states have shown 
the extremely anisotropic longitudinal resistivities and the un-quantized Hall
resistivities at tens of milli Kelvin. 
It is difficult to explain the experiments with the ACDW state. 
On the other hand, 
the HF energy of
the ACDW state is slightly lower than that of the striped Hall state at
each LL, 
as seen in Tables \ref{table:stripe_table} and
\ref{table:ACDW_table}. 
This was one of the remaining issues for the anisotropic states. 
In Sec. \ref{sec:energy}, 
we study the total energy of the present two HF states in the system
with injected currents to answer this issue.

\section{Response functions}
\label{sec:response}
In this section, the electromagnetic response functions 
of the HF states are calculated 
in the long wavelength limit. 
We consider the quantum Hall system with the 
infinitesimal external gauge field 
$a_\mu(x)=(a_0(x), -{\bm a}(x))$ 
and calculate the response
functions of the striped Hall state first and the ACDW state next. 

\subsection{Response function of the striped Hall state}
The Hamiltonian in the quantum Hall system with $a_\mu(x)$ is given by 
\begin{align}
 H=&\int d^2x \Psi^\dag(x)
  \left(
   \frac{({\bf p}+e{\bf A}(x)+e{\bm a}(x))^2}{2m}-ea_0(x) 
  \right)\Psi(x)+\frac{1}{2}\int d^2x d^2x':\rho(x)V(x-x')\rho(x'):, 
\label{eq:Hamiltonian_with_a}
\end{align} 
where $V(x)=q^2/|{\bf x}|$. 
We project the Coulomb interaction part to each LL 
and apply the HF approximation to 
the projected Coulomb interaction. 
Then, using the vNL base, the Hamiltonian 
in the HF approximation is given by
%
\begin{widetext}
\begin{align}
 H=&
 \sum_l E_l\int_{\rm BZ}
 \frac{d^2 p}{(2\pi)^2}b_l^\dag({\bf p})b_l({\bf p})
 \nonumber \\
 &-\int\frac{d^2k}{(2\pi)^2}\sum_{l,l'}
 ef^\mu_{l,l'}(\tilde{\bf k})a_\mu(\tilde{\bf k})
 \int_{\rm BZ}\frac{d^2p}{(2\pi)^2}
 b^\dag_l({\bf p})b_{l'}({\bf p}-{\bf k})
 e^{-(i/4\pi)k_x(2p_y-k_y)}\nonumber\\
 &+\int\frac{d^2k d^2k'}{(2\pi)^4}\sum_{l,l'}
 \frac{e^2\omega_c}{4\pi}
 {\bm a}(\tilde{\bf k})\cdot{\bm a}(\tilde{\bf k}')
 f^0_{l,l'}(\tilde{\bf k}+\tilde{\bf k}')\int_{\rm BZ}\frac{d^2p}{(2\pi)^2}b^\dag_l({\bf p})
 b_{l'}({\bf p}-{\bf k}-{\bf k}')
 e^{-(i/4\pi)(k_x+k'_x)(2p_y-k_y-k'_y)}\nonumber \\
 &+\sum_{l}\int\frac{d^2k}{(2\pi)^2}v_l^{\rm HF}({\bf k})
 \langle \bar{\rho}_l(-\tilde{\bf k})\rangle \bar{\rho}_l(\tilde{\bf k}), 
\end{align}
\end{widetext}
where  
$f^\mu_{l_1,l_2}({\bf k})$ is defined by 
(see Appendix \ref{app:LL_matrix}) 
\begin{equation} 
 f^\mu_{l_1,l_2}({\bf k})=
  \langle f_{l_1}|\frac{1}{2}\{ 
  v^\mu,e^{i{\bf k}\cdot{\bm \xi}} \}|f_{l_2}\rangle
  \label{eq:def_of_f}
\end{equation}
in which $v^\mu=(1,-\omega_c\eta,\omega_c\xi)$ is the electron
velocity. 
Repeated Greek indices $\mu$ and $\nu$ are summed in this paper. 
The action is given by
\begin{align}
 S[a,b,b^\dag]=\int dt 
  \Big[&\int_{\rm
  BZ}\frac{d^2p}{(2\pi)^2}b^\dag_l({\bf p},t)
  (i\partial_t+\mu_{\rm F})b_l({\bf p},t)-H(t)\Big], 
\label{eq:HF_action}
\end{align}
where $H(t)$ is the Heisenberg representation of $H$. 

Let us concentrate on the striped Hall state 
at half-filling. 
Substituting Eq. (\ref{eq:assumption_stripe}) into
Eq. (\ref{eq:HF_action}),  
the action of the striped Hall state is given by
\begin{widetext}
\begin{equation}
S_{\rm HF}[a,b,b^\dag]=\sum_{l,l'}\int_{\rm BZ}\frac{d^3p d^3p'}{(2\pi)^6}
b_l^\dag(p)\Bigl[
(p_0-\xi_l({\bf p}))\delta_{l,l'}(2\pi)^3\delta^3(p-p')
-U_{a_1}^{(l,l')}(p,p')-U_{a_2}^{(l,l')}(p,p')\Bigr]b_{l'}(p'), 
\end{equation}
where
\begin{align*}
 U_{a_1}^{(l,l')}(p,p')=&-\sum_{\bf N}
 ef^\mu_{l,l'}(\tilde{\bf p}-\tilde{\bf p}'-2\pi\tilde{\bf N})
 h({\bf p}+{\bf p}',{\bf N})
 a_\mu(\tilde{\bf p}-\tilde{\bf p}'-2\pi\tilde{\bf N}, p_0-p'_0)e^{-(i/4\pi)(p_x-p'_x)(p_y+p'_y)},
 \nonumber  \\
 U_{a_2}^{(l,l')}(p,p')=&
 \sum_{\bf N}\int\frac{d^3k}{(2\pi)^3}\frac{e^2\omega_c}{4\pi}
 f^0_{l,l'}(\tilde{\bf p}-\tilde{\bf p}'-2\pi\tilde{\bf N})
 h({\bf p}+{\bf p}',{\bf N})\nonumber \\
 &\times 
 {\bm a}(\tilde{\bf k},k_0)
 \cdot {\bm a}(\tilde{\bf p}-\tilde{\bf p}'-\tilde{\bf
 k}-2\pi\tilde{{\bf N}}, p_0-p'_0-k_0)e^{-(i/4\pi)(p_x-p'_x)(p_y+p'_y)}
\end{align*}
\begin{equation}
 h({\bf p},{\bf N})\equiv (-1)^{N_x+N_y+N_xN_y}
 e^{-(i/2)p_xN_y+(i/2)p_yN_x}. 
\end{equation}
\end{widetext}
Here, $p$ denotes $({\bf p}, p_0)$, 
$\xi_l({\bf p})=E_l+\epsilon_l({\bf p})-\mu_{\rm F}$, 
and $U_{a_1}$ and $U_{a_2}$ are the first order term and the second
order term with respect to $a_\mu$, respectively. 

\begin{figure}
 \includegraphics[width=7cm]{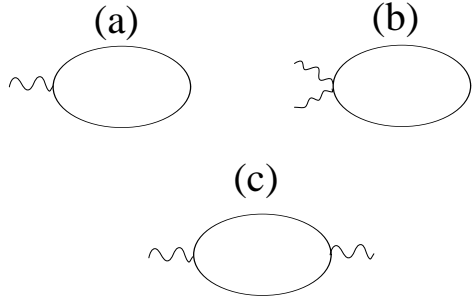}
 \caption{\label{fig:Feynman}(a), (b) and (c) are 
Feynman diagrams for 
$\Delta S_1$, $\Delta S_2$ and $\Delta S_3$, respectively.}
\end{figure}

The partition function $Z[a]$ is calculated using path integrals by
\begin{align}
 Z[a]=&\int\mathcal{D}b^\dag\mathcal{D}b\, 
 e^{iS_{\rm HF}[a,b,b^\dag]}\nonumber \\
 =&\int\mathcal{D}b^\dag\mathcal{D}b\, 
 e^{-(-i)\sum b^\dag(g^{-1}{\bf 1}-U_{a_1}-U_{a_2})b}\nonumber\\
 =&e^{{\rm Tr}\log[(-i)g^{-1}]}
 e^{{\rm Tr}\log[{\bf 1}-gU_{a_1}-gU_{a_2}]},
\end{align}
where 
the power of the exponent is expressed in the matrix
representation in the momentum space and ${\rm Tr}$ denotes the trace of 
the momentum indices and the LL indices. 
$g_l$ is the Green's function given by
\begin{equation}
 g_l(p)=\frac{\theta(\xi_l({\bf p}))}{p_0-\xi_l({\bf p})+i\delta}
  +\frac{\theta(-\xi_l({\bf p}))}{p_0-\xi_l({\bf p})-i\delta}, 
\end{equation}
where $\delta$ is an infinitesimal positive constant. 
The effective action $S_{\rm eff}$ is defined as 
$S_{\rm eff}[a]=-i\log Z[a]$. It consists of 
the non-perturbed part $S_0={\rm Tr}\log[(-i)g^{-1}]$ and 
the correction part due to the external gauge field 
$\Delta S_{\rm eff}[a]$. $\Delta S_{\rm eff}[a]$ is given by 
(Fig. \ref{fig:Feynman}), 
\begin{equation}
 \Delta S_{\rm eff}[a]=
  \Delta S_1[a] +\Delta S_2[a] +\Delta S_3[a] +\mathcal{O}[a^3],
  \label{eq:Seff_stripe}
\end{equation}
where
\begin{widetext}
\begin{align}
 \Delta S_1[a]=&i{\rm Tr}[gU_{a_1}]=
 i\sum_l\int_{\rm BZ}
 \frac{d^3p}{(2\pi)^3}g_l(p)U^{(l,l)}_{a_1}(p,p),\nonumber \\
 \Delta S_2[a]=&i{\rm Tr}[gU_{a_2}]=
 i\sum_l\int_{\rm BZ}
 \frac{d^3p}{(2\pi)^3}g_l(p)U^{(l,l)}_{a_2}(p,p),\nonumber \\
 \Delta S_3[a]=&\frac{i}{2}{\rm Tr}[gU_{a_1}gU_{a_1}]
 =\frac{i}{2}\sum_{l,l'}
 \int_{\rm BZ}\frac{d^3pd^3p'}{(2\pi)^6}
 g_l(p)U^{(l,l')}_{a_1}(p,p')
 g_{l'}(p')U^{(l',l)}_{a_1}(p',p).
 \label{eq:S_stripe}
\end{align}
Substituting the expressions for $g, U_{a1}$ and $U_{a2}$ into
Eq.(\ref{eq:S_stripe}), 
$\Delta S_{\rm eff}[a]$ is given by 
\begin{align}
 \Delta S_{\rm eff}[a]=
 &(e l_0)a_0(0)+\sum_{N_x} e
 f^\mu_{l_0,l_0}(-2\pi\tilde{N}_x,0)\frac{\sin(p_FN_x)}{\pi N_x}
 e^{i\pi N_x}
 a_\mu(-2\pi\tilde{N}_x,0)\nonumber \\
 &-\frac{1}{2}\int\frac{d^3 p}{(2\pi)^3}\sum_{\bf N}
 a_\mu({\bf p},p_0)K^{\mu\nu}(p,{\bf N})a_\nu(-{\bf p}-2\pi\tilde{\bf N},-p_0),
\end{align}
where $l_0$ represents the uppermost partially filled LL. 
$K^{\mu\nu}(p,{\bf N})$ is a response function given by 
\begin{align}
 \label{eq:response_stripe}
 K^{\mu\nu}(p,{\bf N})=&\sum_{l,l'}e^2f^\mu_{l,l'}({\bf p})
  f^\nu_{l',l}(-{\bf p}-2\pi\tilde{\bf N})I_{l,l'}(p_0,\hat{\bf p},{\bf N})
  h(\hat{\bf p}, {\bf N})\nonumber \\
 &+\frac{e^2}{2\pi}\omega_c\sum_{\bf N}\left[
 l_0\delta_{{\bf N},0}+f_{l_0,l_0}^0(-2\pi\tilde{\bf N})(-1)^{N_x}
 \frac{\sin(\pi N_x/2)}{\pi N_x}\delta_{N_y,0}\right]
 (\delta_{\mu,1}\delta_{\nu,1}+\delta_{\mu,2}\delta_{\nu,2}), 
\end{align}
and $I_{l,l'}(p_0,{\bf p},{\bf N})$ is the loop integral in Fig. \ref{fig:Feynman}(c), 
which is given by
\begin{align}
 I_{l,l'}(p_0,{\bf p},{\bf N})=&\int_{\rm BZ}\frac{d^2p'}{(2\pi)^2}
  \Biggl[
   \frac{\theta(\xi_l({\bf p}+{\bf p}'))\theta(-\xi_{l'}({\bf p}'))}
   {p_0+\xi_{l'}({\bf p}')-\xi_l({\bf p}+{\bf p}')+i\delta}
   -\frac{\theta(-\xi_{l}({\bf p}+{\bf p}'))\theta(\xi_{l'}({\bf p}'))}
   {p_0+\xi_{l'}({\bf p}')-\xi_l({\bf p}+{\bf p}')-i\delta}
  \Biggr]e^{-ip'_xN_y+ip'_yN_x}, 
\end{align}
\end{widetext}
in which the $p'_0$ integral has been performed. 
In Eq. (\ref{eq:response_stripe}), 
the first term and the second term  come from 
$\Delta S_3[a]$ and $\Delta S_2[a]$, respectively, 
and 
the second term is cancelled with the $p=0$ part of the first term, as
expected from gauge invariance. 
Hence, $K^{i,i}(p=0,{\bf N})=0$ for $i=1,2$. 

In the long wavelength limit, 
the largest contribution in the response function 
comes from the ${\bf N}=0$ part. 
In the case of $p_0=p_y=0$ and $p_x\to 0$, which is used in the next
section, 
the largest contribution comes from the lowest order term
in $K^{\mu\nu}_0(p_x)\equiv K^{\mu\nu}({p_x,0})$ 
with respect to $p_x$.
By expanding $K^{\mu\nu}_0(p_x)$ up to the lowest order, 
the response functions in the long wavelength limit 
are given as 
\begin{align}
 K^{00}_0(p_x)=&
 -\frac{\sigma^{(\nu)}_{xy}}{\omega_c}p^2_x,\nonumber \\
 K^{0y}_0(p_x)=&-i\sigma^{(\nu)}_{xy}p_x,\nonumber \\
 K^{y0}_0(p_x)=&i\sigma^{(\nu)}_{xy}p_x,\nonumber \\
 K^{yy}_0(p_x)=&\alpha_K\omega_c p^2_x,
\end{align}
where $\sigma^{(\nu)}_{xy}=e^2\nu/2\pi$ 
and $\alpha_K=e^2\omega_c(l_0^2+2l_0\nu^\ast+\nu^\ast)/4\pi^2$. 
$\sigma_{xy}^{(\nu)}$ is identified as the Hall conductance since 
if we consider a static homogeneous electric field in the
$x$-direction generated by the gauge field $a_0^{\rm ex}(x)=xE_x$, then
the electric current in the $y$-direction $j^y(x)$ is given in the long
wavelength limit by 
$\langle j_y(x) \rangle=\delta \Delta 
S_{\rm eff} / \delta a_y(x)=
K^{y0}_0(\partial_x)a_0(x)=-\sigma_{xy}^{(\nu)}E_x$, 
where the response
function transformed in the coordinate space is used.
The longitudinal resistivity becomes zero in the present calculation 
since the impurity potential 
is not included. 
If impurities are added, it is expected that 
the longitudinal resistivity becomes zero in
one direction and finite in the other
direction due to the anisotropic Fermi surface. 

\subsection{Response function of the anisotropic charge density wave state}
The action of the ACDW state at half-filling 
is given by
\begin{widetext}
\begin{align}
S_{\rm HF}[a,{\bf c},{\bf c}^\dag ]=\sum_{l,l'}\int_{\rm RBZ}\frac{d^3pd^3p'}{(2\pi)^6}
 {\bf c}_l^\dag(p)\Big[G^{-1}_l(p)
 \delta_{l,l'}(2\pi)^3\delta^3(p-p')
 -V^{(l,l')}_{a_1}(p;p')-V^{(l,l')}_{a_2}(p;p')\Big]{\bf c}_{l'}(p'), 
\end{align}
where $V_{a_j}(j=1,2)$ is a $2\times 2$ matrix given by
\begin{align}
 V_{a_j}^{(l,l')}(p,p')=&U^{\dag}(p){\bf U}^{(l,l')}_{a_j}(p,p')U(p),
 \nonumber \\
 {\bf U}_{a_j}^{(l,l')}(p;p')=&
 \left(
  \begin{array}{cc}
   U_{a_j}^{(l,l')}(p;p')& U_{a_j}^{(l,l')}(p;p'_x,p'_y+\pi,p'_0) \\
   U_{a_j}^{(l,l')}(p_x,p_y+\pi,p_0;p') 
    & U_{a_j}^{(l,l')}(p_x,p_y+\pi,p_0;p'_x,p'_y+\pi,p'_0)
  \end{array}
 \right), 
\end{align}
and $G_l$ is a $2\times 2$ matrix Green's function given by
\begin{equation}
 G_{l}(p)=\left\{
	   \begin{array}{ll}
	    \left(
	     \begin{array}{cc}
	      g^+_l(p)& 0\\
	      0 & g^-_l(p)
	     \end{array}
	    \right)
	    &{\rm for} \quad l=l_0\\
	    g^-_l(p){\bf 1}&{\rm for} \quad l<l_0\\
	    g^+_l(p){\bf 1}&{\rm for} \quad l>l_0
	   \end{array}
	   \right. .
\label{eq:Green}
\end{equation}
\end{widetext}
In Eq. (\ref{eq:Green}), 
$g^{\pm}_l(p)=1/(p_0-\xi^\pm_l(p)\pm i\delta)$ is 
a one-particle Green's function for the upper or lower band, 
${\bf 1}$ is a $2\times 2$ unit matrix,
and $\xi^{\pm}_l({\bf p})=E_{l_0}+\epsilon^{(l_0)}_{\pm}({\bf p})-\mu_{\rm F}$. 

The partition function $Z[a]$ is calculated using path integrals by
\begin{align}
 Z[a]=&\int\mathcal{D}{\bf c}^\dag\mathcal{D}{\bf c}\, 
 e^{iS_{\rm HF}[a,{\bf c},{\bf c}^dag]}\nonumber \\
 =&e^{{\rm Tr}\log[(-i)G^{-1}]}
 e^{{\rm Tr}\log[{\bf 1}-GV_{a_1}-GV_{a_2}]}. 
\end{align}
The correction part of the effective action 
$\Delta S_{\rm eff}[a]$ is given by
\begin{equation}
 \Delta S_{\rm eff}[a]=
  \Delta S_1[a] +\Delta S_2[a] +\Delta S_3[a] +\mathcal{O}[a^3],
  \label{eq:Seff_ACDW}
\end{equation}
where
\begin{widetext}
\begin{align}
 \Delta S_1[a]=&i{\rm Tr}[GV_{a_1}]
 =
 i\sum_l\int_{\rm RBZ}\frac{d^3p}{(2\pi)^3}
 {\rm Tr_{2\times 2}}[G_l(p)V^{(l,l)}_{a_1}(p,p)],
 \nonumber \\
 \Delta S_2[a]=&i{\rm Tr}[GV_{a_2}]
 =
 i\sum_l\int_{\rm RBZ}\frac{d^3p}{(2\pi)^3}
 {\rm Tr_{2\times 2}}[G_l(p)V^{(l,l)}_{a_2}(p,p)],
 \nonumber \\
 \Delta S_3[a]=&\frac{i}{2}{\rm Tr}[GV_{a_1}GV_{a_1}]
 =\frac{i}{2}\sum_{l,l'}
 \int_{\rm RBZ}\frac{d^3pd^3p'}{(2\pi)^6}
 {\rm Tr_{2\times 2}}
 [G_l(p)V^{(l,l')}_{a_1}(p,p')
 G_{l'}(p')V^{(l',l)}_{a_1}(p',p)]. 
 \label{eq:Delta_ACDW}
\end{align}
Here, ${\rm Tr}$ denotes the trace with respect to the momentum
indices, the LL indices and
the $2\times 2$ matrix indices, 
and ${\rm Tr}_{2\times 2}$ denotes the
trace with respect to only the $2\times 2$ matrix indices. 
Substituting the expressions for $G,V_{a_1},V_{a_2}$ into
Eq.(\ref{eq:Delta_ACDW}),
we obtain the following expression for 
the ${\bf N}=0$ part of $\Delta S_{\rm eff}$ 
as in the case of the striped Hall state:
\begin{align}
 \Delta S_{\rm eff}[a]=&
 e(l_0+\frac{1}{2}) a_0(0)
 -\frac{1}{2}\int\frac{d^3p}{(2\pi)^3}
 a_{\mu}({\bf p},p_0)K_0^{\mu\nu}(p)a_\nu(-{\bf p},-p_0), 
\end{align}
where $K^{\mu\nu}_0$ is given by
\begin{align}
 \label{eq:response_ACDW}
 K^{\mu\nu}_0(p)=&
 -\sum_{l<l_0}\sum_{l'>l_0}
 \frac{e^2}{\omega_c(l'-l)}
 \left\{f^\mu_{l,l'}({\bf p})f^\nu_{l',l}(-{\bf p})
 +f^\mu_{l',l}({\bf p})f^\nu_{l,l'}(-{\bf p}))\right\}
 \nonumber \\
 &+\frac{1}{2}
 \left[-\sum_{l<l_0}+\sum_{l>l_0}\right]
 \frac{e^2}{\omega_c(l_0-l)}
 \left\{f^\mu_{l,l_0}({\bf p})f^\nu_{l_0,l}(-{\bf p})
 +f^\mu_{l_0,l}({\bf p})f^\nu_{l,l_0}(-{\bf p})
 \right\}
 \nonumber \\
 &+\int_{\rm RBZ}\frac{d^2p'}{(2\pi)^2} 
 \frac{1}{p_0-(\epsilon({\bf p}')+\epsilon(\hat{\bf p}+{\bf p}'))}
 \nonumber \\
 &\qquad \times 
 \left[1-
 \frac{A(\hat{\bf p}+{\bf p}')A({\bf p}')
 +{\rm Re}(B(\hat{\bf p}+{\bf p}')B^\ast({\bf p}')
 e^{-(i/2)\hat{p}_x})}
 {\epsilon(\hat{\bf p}+{\bf p}')\epsilon({\bf p}')}\right]
 e^2f^\mu_{l_0,l_0}({\bf p})f^\nu_{l_0,l_0}(-{\bf p})
 \nonumber \\
 &+(l_0+\frac{1}{2})\frac{e^2}{2\pi}\omega_c
 (\delta_{\mu,1}\delta_{\nu,1}+\delta_{\mu,2}\delta_{\nu,2}), 
\end{align}
\end{widetext}
up to $\mathcal{O}(\epsilon({\bf p})/\omega_C)$. 
In Eq. (\ref{eq:response_ACDW}), the last term is cancelled with 
the $p=0$ term of the second term 
as expected from gauge invariance
again. Hence, $K_0^{i,i}(0)=0$ for $i=1,2$. 

In the long wavelength limit $p_0=p_y=0$ and $p_x\to 0$, 
the largest contribution in the response function 
comes from the lowest order term with respect to $p_x$. 
The expressions of $K_0^{0y}$ and $K_0^{y0}$ become the same as in
the case of the striped Hall state. 
The expression of $K_0^{00}$ becomes slightly different by the 
correction from the intra-LL effect 
at the uppermost partially-filled LL.
For the striped Hall state, 
the one-particle energy shown in Fig. \ref{fig:ene_stripeL2} has the 
inter-LL energy gap in the $p_x$-direction, and 
the inter-LL effect gives the response functions given 
in Eq. (\ref{eq:response_stripe}). 
For the ACDW state, 
the one-particle energy shown in Fig. \ref{fig:ene_ACDWL2} has the 
intra-LL energy gap in the $p_x$-direction as well as the inter-LL
energy gap. 
While the inter-LL effect gives the same expression of the response function
as that of the striped Hall state, 
the intra-LL effect causes some corrections to the response function. 
Including these corrections, $K_0^{00}$ 
of the ACDW state in the long wavelength limit is given by (see
Appendix \ref{app:K00}) 
\begin{equation}
 K^{00}_0(p_x)=-(1+\frac{\sqrt{B}}{\nu}\beta)
  \frac{\sigma_{xy}^{(\nu)}}{\omega_c}p_x^2, 
  \label{eq:K_for_ACDW}
\end{equation}
where $a=\sqrt{2\pi/eB}$ is used explicitly in order to compare the
theoretical results with experimental data. 
The value of $\beta$ at each LL 
is shown in Table \ref{table:beta}. 
Note that the unit of $\beta$ is $({\rm tesla})^{-1/2}$. 
\begin{table} 
\caption{\label{table:beta}The value of $\beta$ at each LL. {\it parallel} is the 
case in which the ACDW direction faces to the $y$-direction. 
{\it perpendicular} is the case in which the ACDW direction faces to the
 $x$-direction. The unit of $\beta$ is $({\rm tesla})^{-1/2}$.}
\begin{ruledtabular}
\begin{tabular}{ccc}
$l$ & parallel  & perpendicular \\
\colrule
0 & 0.497  & 0.497 \\
1 & 0.386  & 0.824 \\
2 & 0.472  & 1.044 \\
3 & 0.581  & 1.171 \\
\end{tabular}
\end{ruledtabular}
\end{table}
The Hall conductance is given by $\sigma_{xy}^{(\nu)}$, as in the case of
the striped Hall state. 
The longitudinal resistivity becomes zero in the present calculation 
since the impurity potential 
is not included. 
However it is expected that 
the longitudinal resistivity remains zero even in the system with
impurities because of the energy gaps.

 
\section{Energy corrections due to finite electric currents}
\label{sec:energy}
In this section, we consider the quantum Hall system with an injected
electric current 
and 
investigate the current effect on the striped Hall state and the ACDW
state. 
Effects of impurities and metallic contacts are ignored in our
calculations.
For the striped Hall state, we only consider the current parallel to the
stripe direction since in this case, the current effect can be estimated
with no ambiguity even in the system with impurities. 
When the current flows in the stripe direction, 
charges accumulate around both edges of the sample in the perpendicular
direction, as we will see later, 
and the electric field generates in the perpendicular direction. 
In this case, the impurity effect is negligible
since the Fermi surface has the inter-LL energy gap 
in the perpendicular direction. 
On the other hand, 
when the current flows in the perpendicular direction, 
the impurity effect becomes relevant since 
the electric field generates in the stripe direction 
while the Fermi surface is gapless in this direction. 
The current effect in this case is nontrivial and will be studied 
in future work. 
 
In the system with an injected current, 
it is naively expected that the current flow causes the plus and minus 
charge accumulation at both edges of the sample with the opposite sign, 
as expected from the classical Hall effect. 
MacDonald et al. have studied the injected current effect on the 
integer quantum Hall system about two decades ago.\cite{Mac1} 
They have calculated the current and charge distributions 
and found that the charge accumulation occurs in the integer
quantum Hall system. 
The charge accumulation causes 
the energy correction via the Coulomb interaction between the accumulated
charges. 
It is expected that the same type of the energy correction exists even 
in the present highly correlated quantum Hall states. 
However it has not been studied as far as the present authors know. 
In the following, 
we first derive the current and charge distributions in the striped
Hall state and the ACDW state using the effective action, 
then, we estimate the current dependence of the energy corrections of 
the two HF states. 
It is shown that the energy of the ACDW state increases faster than 
that of the striped Hall state as the injected current increases. 

\subsection{Current and charge distributions}
We study current and charge distributions of the striped Hall state and
the ACDW state. 
We denote the two-point function in the HF theory 
 with no injected current 
$\langle \Psi^\dag({\bf x}, t)\Psi({\bf x}', t)\rangle_{I=0}$ 
as $F({\bf x},{\bf x}')$ for both states. 
In the system with a finite electric current, 
electromagnetic fields and the two-point function 
deviate from their original values. 
These deviations are taken into account 
in the calculation of the total energy. 
We define these deviations by 
\begin{align}
 \label{eq:def_of_deviations}
 {\bm a}({\bf x}, t)=&{\bf A}({\bf x},t)-{\bf A}_{\rm ex}({\bf x}), \nonumber \\
 \delta\rho({\bf x},{\bf x}',t)=&
 \langle \Psi^\dag({\bf x},t)\Psi({\bf x}',t)\rangle - F({\bf x},{\bf x}'), 
\end{align}
where ${\bm a}$ and $\delta\rho$ are unspecified for the moment and 
will be determined later. 
The total action in the Coulomb gauge 
$\nabla \cdot {\bf A}(x)=0$ is given as 
\begin{widetext}
\begin{align}
 S_{\rm tot}[{\bf A}, \Psi^\dag, \Psi]=&\int dtd^3x
 \left( \frac{\epsilon}{2}{\dot {\bf A}}^2({\bf x},t)-\frac{1}{2\mu}
 (\nabla\times{\bf A}({\bf x},t))^2 \right)\nonumber \\
 &+\int dtd^3x \Psi^\dag({\bf x},t)
 \left(i\partial_t-\frac{({\bf p}+e{\bf
 A}({\bf x},t))^2}{2m}\right)\Psi({\bf x},t)\delta(z)\nonumber \\
 &-\frac{1}{2}\int dt d^3x d^3x'
 \Psi^\dag({\bf x},t)\Psi^\dag({\bf x}',t)V({\bf x}-{\bf x}')
 \Psi({\bf x}',t)\Psi({\bf x},t)\delta(z)\delta(z').
\end{align}
\end{widetext}
where $\mu$ is the magnetic constant and the dot means the
time derivative. 
This total action consists of 
the three-dimensional electromagnetic field term and the two-dimensional 
electron field term. 
In the Coulomb gauge, the 
interaction between electric fields is expressed by the Coulomb
interaction. 
Applying the HF approximation to the Coulomb interaction part 
and substituting Eq. (\ref{eq:def_of_deviations}), 
the total action is rewritten as 
\begin{equation}
 S_{\rm tot}[{\bm a},\delta\rho,\Psi^\dag,\Psi]=S_{\rm EM}[{\bm a}]
  +S_{\rm HF}[{\bm a},\delta\rho,\Psi^\dag,\Psi],
\end{equation}
where
\begin{widetext}
\begin{align}
 S_{\rm EM}[{\bm a}]=&\int dtd^3x
 \left(\frac{\epsilon}{2}\dot{{\bm a}}^2({\bf x},t)
 -\frac{1}{2\mu}(\nabla\times{\bm a}({\bf x},t))^2
 \right), \nonumber \\
 S_{\rm HF}[{\bm a},\delta\rho,\Psi^\dag,\Psi]=
 &\int dtd^3x\Psi^\dag({\bf x},t)
 \left(i\partial_t-\frac{({\bf p}+e{\bf A}_{\rm ex}({\bf x})
 +e{\bm a}({\bf x},t))^2}{2m}\right)\Psi({\bf x},t)\delta(z)\nonumber \\
 &-\int dtd^3xd^3x'\Big[
 (F({\bf x},{\bf x})+\delta\rho({\bf x},{\bf x},t))
 V({\bf x}-{\bf x'})\Psi^\dag({\bf x}',t)\Psi({\bf x}',t)\nonumber \\
 &-(F({\bf x},{\bf x}')+\delta\rho({\bf x},{\bf x}',t))
 V({\bf x}-{\bf x'})\Psi^\dag({\bf x}',t)\Psi({\bf x},t)\Big]\delta(z)\delta(z').
\label{eq:total_action}
\end{align}
\end{widetext}
In the expression of $S_{\rm EM}$, 
the term of the uniform external magnetic field is dropped
since it gives only the same energy constant to the two HF states. 
In Eq. (\ref{eq:total_action}), the term including 
$\delta\rho({\bf x,}{\bf x},t)$ 
and the term including $\delta\rho({\bf x},{\bf x}',t)$ are 
the Hartree term and the Fock term, respectively. 
As seen in Appendix \ref{app:HF_Hamiltonian}, 
the Fock term becomes negligible compared to the Hartree term 
in the long wavelength limit 
since in the momentum space, 
the Hartree term is proportional to the Coulomb potential $V({\bf k})$, 
which is $\mathcal{O}(1/k)$, and gives a larger contribution than the Fock
term for 
the small momentum $k$.
In the following calculation, the deviation of
the Fock term is dropped. 
If we introduce the potential generated by the electron density deviation as
\begin{equation}
 a_0({\bf x},t)\equiv \int d^3x'
  \frac{(-e) \delta\rho({\bf x}',{\bf x}',t)}
  {4\pi\epsilon|{\bf x}-{\bf x}'|}
  \delta(z').
  \label{eq:scalar_potential}
\end{equation}
$S_{\rm HF}$ is rewritten as
\begin{widetext}
\begin{align}
 S_{\rm HF}[{\bm a},a_0,\Psi^\dag,\Psi]=&\int dtd^3x\Psi^\dag({\bf x},t)
 \left(i\partial_t+e a_0({\bf x},t)-
 \frac{({\bf p}+e {\bf A}_{\rm ex}({\bf x})
 +e {\bm a}({\bf x},t))^2}{2m}\right)\Psi({\bf x},t)\delta(z)\nonumber \\
 &-\int dtd^3xd^3x'\Big[
 F({\bf x},{\bf x}) V({\bf x}-{\bf x'})\Psi^\dag({\bf x}',t)\Psi({\bf x}',t)\nonumber \\
 &\qquad -F({\bf x},{\bf x}')V({\bf x}-{\bf
 x'})\Psi^\dag({\bf x}',t)\Psi({\bf x},t)\Big]\delta(z)\delta(z'). 
\end{align}
\end{widetext}
The same form of the action is obtained from the  
Hamiltonian in the system with the infinitesimal external 
gauge field shown in Eq. (\ref{eq:Hamiltonian_with_a}) when the 
Coulomb interaction part is approximated in the 
HF approximation. 
The important difference is that 
$a_\mu$ in the present case represents the 
finite gauge field induced by the current flow. 
Although the meaning of $a_\mu$ is different, 
the effective action obtained in the previous section is 
applicable as long as $a_\mu$ is small. 

The partition function is given by 
\begin{equation}
 Z=\int {\mathcal D}{\bm a}\int {\mathcal D} 
  \Psi^\dag {\mathcal D}\Psi\, 
  e^{iS_{\rm EM}[{\bm a}]+iS_{\rm HF}[{\bm a},a_0,\Psi^\dag,\Psi]}. 
\end{equation}
Integrating out electron fields and expanding the results 
up to second order of ${\bm a}$ and $a_0$, 
we obtain the effective action $S_{\rm eff}$ as
\begin{equation}
 Z=\int {\mathcal D}{\bm a}\, e^{iS_{\rm EM}[{\bm a}]+
  iS_0+i\Delta S_{\rm eff}[{\bm a},a_0]}.
\end{equation}
The functional derivative of $(S_{\rm EM}+S_0+\Delta S_{\rm eff})$ with respect to
${\bm a}({\bf x}, t)$ gives the Maxwell's equation for ${\bm a}({\bf
x},t)$: 
\begin{equation}
 (\epsilon \partial^2_t-\frac{1}{\mu}\nabla^2){\bm a}({\bf x},t)
  =\langle {\bf j}({\bf x},t) \rangle_a\delta(z),
  \label{eq:vector_potential}
\end{equation}
where ${\bf j}({\bf x},t)$ is a current operator and 
$\langle \hat{O}(x) \rangle_a$ means an expectation value of an operator
$\hat{O}(x)$ for the system with finite $a_\mu$. 
The solution of this equation gives the stationary point of the action 
with respect to $a_\mu$. 
We use the action 
into which the solution of Eq. (\ref{eq:vector_potential}) is
substituted as the effective action. 
$\langle {\bf j}({\bf x},t) \rangle_a$ and 
$\delta\rho({\bf x},t) \equiv \delta\rho({\bf x},{\bf x},t)$ are
calculated from the effective action by
\begin{align}
 \frac{\delta \Delta S_{\rm eff}
 [{\bm a},a_0]}{\delta{\bm a}({\bf x},t)}=&\langle {\bf j}({\bf x},t)
 \rangle_a\delta(z), \nonumber \\
 -\frac{\delta \Delta S_{\rm eff}[{\bm a},a_0]}{\delta a_0({\bf x},t)}=&
 (-e)[\rho_0({\bf x})+\delta\rho({\bf x},t)]\delta(z),
 \label{eq:eff_to_j}
\end{align}
where the $\rho_0({\bf x})$ is the expectation value of the density operator in
the system with no injected current. Eq.(\ref{eq:scalar_potential}), 
(\ref{eq:vector_potential}) and (\ref{eq:eff_to_j}) determine 
$a_0({\bf x},t)$ and ${\bm a}({\bf x},t)$, 
or $\delta\rho({\bf x},t)$ and $\langle {\bf j}({\bf x},t) \rangle_a$, self-consistently.

\begin{figure}
\includegraphics[width=5cm]{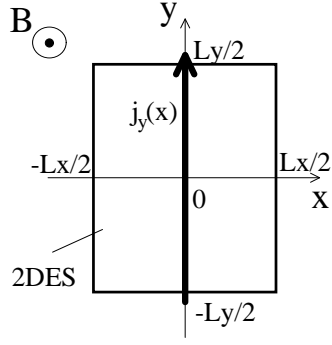}
\caption{\label{fig:Current}
Schematic view of the 2D electron system in a magnetic field 
with the injected current. 
The current flows in the $y$-direction and has only 
the $x$-coordinate dependence.
}
\end{figure}

We concentrate on the finite system with  
the static injected current flowing in the
$y$-direction and depending only on $x$. 
The lengths of the 2D electron system in the $x$-direction and the $y$-direction are 
$L_x$ and $L_y$, respectively (Fig. \ref{fig:Current}). 
In this case, the electron density also depends only on $x$ and 
Eq.(\ref{eq:scalar_potential}) and Eq.(\ref{eq:vector_potential})
gives the following solutions at $z=0$: 
\begin{align}
 a_0(x)=&-\frac{1}{2\pi\epsilon}\int^{L_x/2}_{-L_x/2} dx'\ln|x-x'|
 (-e)\delta\rho(x')\nonumber ,\\
 a_y(x)=&\frac{\mu}{2\pi}\int^{L_x/2}_{-L_x/2} dx'\ln|x-x'|
 \langle j_y(x') \rangle _a.
 \label{eq:self_eq1}
\end{align} 
As shown in the previous section, 
the effective action can be divided into 
the non-perturbed part and the correction part due to currents,
and in the long wavelength limit, 
the correction part $\Delta S_{\rm eff}$ is given as
\begin{align}
 \Delta S_{\rm eff}[{\bm a},a_0]=&
  -TL_y\int^{L_x/2}_{-L_x/2} dx (-e)\bar{\rho}_0 a_0(x)-
  \frac{TL_y}{2}\int^{L_x/2}_{-L_x/2} dx\, 
  a_\mu(x)K^{\mu\nu}_0(\partial_x) a_\nu(x),
  \label{eq:Delta_S_eff}
\end{align}
where $\bar{\rho}_0$ is a uniform part of the density, and 
$T$ is the total time. 
$K_0^{\mu\nu}(\partial_x)$ is the Fourier transformed form 
of the response function obtained in the previous section. 
Substituting Eq. (\ref{eq:Delta_S_eff}) into Eq.({\ref{eq:eff_to_j}}), 
$\delta\rho(x)$ and $\langle j_y(x) \rangle_a$ are
given as
\begin{align}
 (-e)\delta\rho(x)=&K^{00}_0(\partial_x)a_0(x)
 -K^{0y}_0({\partial_x})a_y(x)\nonumber, \\
 \langle j_y(x)\rangle_a=&
 K^{y0}_0(\partial_x)a_0(x)-K^{yy}_0(\partial_x)a_y(x).
 \label{eq:self_eq2}
\end{align}
Equations (\ref{eq:self_eq1}) and ({\ref{eq:self_eq2}}) determine
the current and charge density distributions up to an overall constant. 
The overall constant is determined by requiring 
the following constraints: 
\begin{equation}
 \int^{L_x/2}_{-L_x/2}dxj_y(x)=I,\qquad \int^{L_x/2}_{-L_x/2}dx\delta\rho(x)=0, 
\end{equation}
where $I$ is a total current. 
Using the explicit form of the response functions derived 
in Sec. \ref{sec:response}, 
we obtain the integral equations to determine 
the current and charge distribution.
 
The same type of the integral equations has already been solved for the 
integer quantum Hall state.\cite{Mac1,Mac2,Mac3,Mac4} 
Their results are summarized as follows: 
(i) 
In Eq. (\ref{eq:self_eq2}), the terms including the vector potential
$a_y(x)$ give a very small effect in the integral equations 
compared to the terms including the scalar potential $a_0(x)$ 
and the vector potential terms are negligible in a good approximation, 
(ii) the analytical solution of the integral equation without the vector
potential term 
is obtained by means of the Wiener-Hopf technique, and 
(iii) $a_0(x)={\rm const}\times \ln|(x-L_x/2)/(x+L_x/2)|$ is the good
approximate form of the analytical solution except near the edge 
and the constant coefficient is determined 
from the constraint for the total current. 
The same results hold in our case. 

\begin{figure}
\includegraphics[width=7cm]{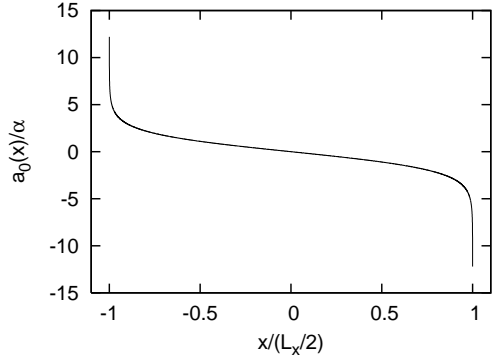}
\caption{\label{fig:potential}Potential distribution $a_0(x)$. 
The first derivative of $a_0(x)$ gives 
the current distribution and the second derivative gives 
the charge distribution.}
\end{figure}

The integral equation for the potential is given as 
\begin{equation}
 a_0(x)=-\gamma
  \int^{L_x/2}_{-L_x/2}dx'\ln|x-x'|\partial^2_{x'}a_0(x'),
  \label{eq:integral_eq} 
\end{equation}
where $\gamma=(1+\beta \sqrt{B}/\nu)\sigma^{(\nu)}_{xy}/2\pi\epsilon\omega_c$ 
($\beta=0$ for the striped Hall state). 
$\gamma$ has the dimension of length and is very small for the magnetic
fields of the order of 
several tesla in the quantum Hall regime. 
For example, 
if $\epsilon=13\epsilon_0$, $m=0.067m_e$ (these are parameters in GaAs), 
and $\beta=0$, then $\gamma$ is of the order of $10^{-8}$ m. 
The current and charge distributions are obtained from $a_0(x)$ as 
\begin{equation}
 \label{eq:rho}
 (-e)\delta\rho(x)=2\pi\epsilon\gamma
 \partial^2_x a_0(x),\quad 
 \langle j_y(x)\rangle_a=-\sigma^{(\nu)}_{xy}\partial_x a_0(x).
\end{equation}
The approximate solution of Eq.(\ref{eq:integral_eq}) is given by 
(Fig. \ref{fig:potential})
\begin{equation}
 a_0(x)=\alpha\ln\left|\frac{x-L_x/2}{x+L_x/2}\right|\qquad
  {\rm for}\quad |x|\le \frac{L_x}{2}-\gamma,
  \label{eq:a_0}
\end{equation}
with a linear extrapolation of $a_0$ to $\pm IR_H/2$ 
in the interval within $\gamma$ from the edge, 
where $\alpha=IR_H/2(1+\ln(L_x/\gamma))$ and $R_H=1/\sigma^{(\nu)}_{xy}$
is the Hall resistivity. 
One may verify that Eq. (\ref{eq:a_0})
is indeed the approximate solution of the integral Eq. 
(\ref{eq:integral_eq}), by substituting 
Eq. (\ref{eq:a_0}) into Eq. (\ref{eq:integral_eq}) and performing one
partial integration. 

\subsection{Energy corrections}
The energy correction due to the injected current per 
unit space-time volume is calculated from the effective action by 
$(S_{\rm EM}+\Delta S_{\rm eff})/TL_xL_y$. 
Since in the present case of $\nu^\ast=1/2$, 
the area occupied by one particle at the uppermost partially-filled LL is $2a^2$ 
(here, the vNL constant $a$ is written explicitly), 
the energy correction per particle $\delta E$ is given by 
$[(S_{\rm EM}+\Delta S_{\rm eff})/TL_xL_y]\times 2a^2$. 
Substituting Eqs.(\ref{eq:self_eq1}) and ({\ref{eq:self_eq2}}) into
this expression, 
the energy correction per particle is given by 
\begin{equation}
 \delta E[I]=-\frac{e^2}{2\pi\epsilon L_x}
  \int^{L_x/2}_{-L_x/2}dxdx'\delta\rho(x)\ln|x-x'|\delta\rho(x'). 
  \label{eq:energy_correction}
\end{equation}
Substituting Eq.(\ref{eq:rho}) into Eq.(\ref{eq:energy_correction}) and
using Eq. (\ref{eq:integral_eq}), 
the energy correction is written as 
\begin{equation}
 \label{eq:energy_correction2}
 \delta E[I]=\frac{(\sigma_{xy}^{(\nu)})^2}
  {2\pi\epsilon\gamma L_x\omega_c^2}
  \int^{L_x/2}_{-L_x/2}dx\, a_0(x)\partial_x^2a_0(x). 
\end{equation}
The final result is obtained by substituting Eq. (\ref{eq:a_0}) into 
this expression and performing the $x$ integral, 
\begin{equation}
 \delta E[I]=\frac{\pi\epsilon}{L_x(\sigma_{xy}^{(\nu)})^2}\times
  \frac{\ln(2/b)-1}{(\ln(2/b)+1)^2}\times I^2. 
\end{equation}
where $b$ is a dimensionless constant given by $b=\gamma/(L_x/2)(\ll 1)$. 
This expression depends on the filling factor, 
the magnetic field strength and experimental parameters. 
Since the actual filling factor includes the spin degree of freedom, 
we use $\nu_{\rm ex}= 2l_0+\nu^\ast$ for lower spin
bands and $\nu_{\rm ex}= (2l_0+1)+\nu^\ast$ for upper spin bands 
instead of $\nu$. 
The magnetic field strength is related to the filling factor by 
$B=h n_e/e \nu_{\rm ex}$ ($n_e$ is an electron density). 
For example, if $n_e=2.67\times 10^{15} {\rm m}^{-2}$, then the magnetic
field strengths are 4.42 T ($\nu_{\rm ex}=5/2$), 3.15 T ($\nu_{\rm ex}=7/2$), 
2.45 T ($\nu_{\rm ex}=9/2$), 2.01 T ($\nu_{\rm ex}=11/2$), 1.70 T
($\nu_{\rm ex}=13/2$) 
and so on. 
We use $\epsilon=13\epsilon_0$, $m=0.067m_e$, 
$n_e=2.67\times 10^{15}$ m, and $L_x=5\times 10^{-3}$ m in order to estimate the values of
energy corrections, which are the parameters used in the experiment by
Lilly et al.\cite{stripe_ex1} 
Then the energy correction is given by 
$\delta E[I]=C\times I^2 (q^2/l_B)$ with the coefficient $C$ shown
in Table \ref{table:C}. 
\begin{table}
\caption{Values of the coefficient $C$ in units of ${\rm nA}^{-2}$ and the
 critical current $I_C$ in units of nA. 
{\it parallel} is the case in which the ACDW direction 
is parallel to the current. 
{\it perpendicular} is the case in which the ACDW direction is 
perpendicular to the current. 
}
\label{table:C}
\begin{ruledtabular}
\begin{tabular}{ccccc}
$\nu_{\rm ex}$ & stripe & parallel  & perpendicular & $I_c$\\
\colrule
5/2  & 325.0  & 330.6   & 335.6   & 0.041 \\
7/2  & 204.4  & 206.6   & 209.0   & 0.065 \\
9/2  & 144.7  & 146.1   & 147.6   & 0.040 \\
11/2 & 109.9  & 110.7   & 111.6   & 0.053 \\
13/2 & 87.44  & 88.08   & 88.68   & 0.047 \\
\end{tabular}
\end{ruledtabular}
\end{table}

\begin{figure}
\includegraphics[width=7cm]{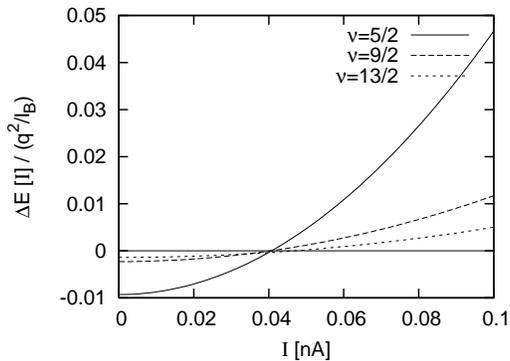}
\caption{\label{fig:ene_diff}Energy differences $\Delta E[I]$
between the striped Hall state 
and the ACDW state. The results at 
$\nu=5/2, 9/2, 13/2$ are shown. 
When $\Delta\epsilon[I]$ is positive, 
the striped Hall state has lower energy.}
\end{figure}

As shown in Sec. \ref{sec:HF}, in the system with no injected current 
the energy of the ACDW state is slightly lower than 
that of the striped Hall state. The differences of energy per particle
$\Delta E_0$ 
are 
$9.3\times 10^{-3}$ ($l_0=1$), $2.3\times 10^{-3}$ ($l_0=2$), 
$1.4\times 10^{-3}$ ($l_0=3$) and so on in units of $q^2/l_B$. 
When the finite current is injected, 
charges are accumulated in both edges with the opposite sign. 
The accumulated charges give 
the energy corrections $\delta E[I]$ which depend on the value of
current $I$. Including these corrections, 
the energy difference between the striped Hall state and the ACDW state 
$\Delta E[I]=-\Delta E_0 + 
(\delta E_{\rm ACDW}[I] - \delta E_{\rm stripe}[I])$ 
varies depending on $I$. 
The current dependence of $\Delta E[I]$ is shown in
Fig. \ref{fig:ene_diff}. 
In Fig .\ref{fig:ene_diff}, 
only the parallel case is plotted for the ACDW states since 
it has a weaker current dependence than the perpendicular case does. 
The signs of the energy differences change 
at the critical values of current $I_c$. 
The critical values are shown in Table \ref{table:C}. 
The critical values are about $0.04$ - $0.07$ nA. 
The current used in the experiments \cite{stripe_ex1,stripe_ex2} 
is above $1$ nA and is much larger than the critical value. 
At $1$ nA, the energy differences $\Delta \epsilon[I]$ become 
$5.6$ $(\nu=5/2)$, $1.4$ $(\nu=9/2)$ and $0.63$ $(\nu=13/2)$ in units of
$q^2/l_B$ 
which are much larger than the original energy differences at a zero
injected current. 
Hence, the striped Hall state becomes the lower energy state and 
should be realized in the experiments. 

\section{Summary}
\label{sec:summary}
In this paper, 
we have investigated the effect of the finite electric current on the 
striped Hall state and the ACDW state 
in the system with no impurities and no metallic contacts 
using the effective action. 
We calculated the electromagnetic response functions and
obtained the effective action. 
For the striped Hall state, 
the current parallel to the stripe direction was investigated. 
In this case, the current effect can be estimated 
with no ambiguity even in the system with impurities. 
The current and charge distributions were 
determined for both states in the system with the injected current.
It is found that the charge accumulation occurs 
around both edges with the opposite sign, 
just as in the case of the integer quantum Hall state studied by 
MacDonald et.al and other authors. \cite{Mac1,Mac2,Mac3,Mac4} 
We hope that current and charge distributions 
will be observed in experiments for anisotropic states. 
The charge accumulation results in the energy enhancement 
via the Coulomb interaction 
between the accumulated charges. 
The energy enhancement was estimated 
from the current and charge distributions. 
It is found that the energy of the ACDW state increases 
faster than that of the striped Hall state does 
as the injected current increases. 
In the system with no injected current, the energy of 
the ACDW state is lower 
than that of the striped Hall state. 
Hence, the striped Hall state becomes the lower energy state 
when the current exceeds the critical value.
The critical value is estimated at about $0.04$ - $0.07$ nA. 
The current used in the experiments for 
the anisotropic states \cite{stripe_ex1,stripe_ex2} is 
above $1$ nA. 
This result suggests that the striped Hall state is realized in the
experiments. 
In addition, 
the striped Hall state has the anisotropic Fermi surface, which 
naively explains 
the experimental features of the anisotropic states, i.e., 
the anisotropic longitudinal resistivities and the un-quantized Hall
resistivities. 
Hence, we conclude that 
the striped Hall state is realized in
the experiment rather than the ACDW state 
and 
predict that the ACDW state is realized if the experiment 
is done with the current smaller than the critical value. 

\begin{acknowledgments}
This work was partially supported by the special Grant-in-Aid for
Promotion of Education and Science in Hokkaido University, 
a Grant-in-Aid for Scientific Research on Priority Area 
(Dynamics of Superstrings and Field Theories, Grant No. 13135201) 
and (Progress in Elementary Particle Physics of the 21st Century through
Discoveries of Higgs Boson and Supersymmetry, Grant No. 16081201), 
provided by the Ministry of Education, Culture, Sports, Science, and 
Technology, Japan. 
\end{acknowledgments}

\appendix

\section{Hartree-Fock Hamiltonian}
\label{app:HF_Hamiltonian}
Let us consider the quantum Hall system and concentrate on the Coulomb interaction part
of the Hamiltonian. 
The Coulomb interaction part is given by 
\begin{equation}
 H_{\rm int}=\frac{1}{2}\int\frac{d^2k}{(2\pi)^2}
  :\rho({\bf k})V({\bf k})\rho(-{\bf k}):, 
\end{equation}
where $\rho({\bf k})$ and $V({\bf k})$ are given in 
Eq. (\ref{eq:expression_of_rho}) and
Eq. (\ref{eq:Coulomb}), respectively. 
The HF approximated form of $H_{\rm int}$ is written using the vNL base 
by 
$H_{\rm HF}=\mathcal{H}_{\rm HF}-\langle \mathcal{H}_{\rm HF}\rangle /2$, 
where  
\begin{gather}
 \mathcal{H}_{\rm HF}=\sum_{l_1,l_2,l_3,l_4}
  \int\frac{d^2k}{(2\pi)^2}v^{\rm HF}_{l_1,l_2,l_3,l_4}(\tilde{\bf k})
  \langle\bar{\rho}_{l_1,l_2}(-\tilde{\bf k})\rangle
  \bar{\rho}_{l_3,l_4}(\tilde{\bf k}), \notag \\
 \bar{\rho}_{l_1,l_2}({\bf k})=\int_{\rm BZ}\frac{d^2p}{(2\pi)^2}
 b^\dag_{l_1}({\bf p})b_{l_2}({\bf p}-\hat{\bf k})
 e^{-(i/4\pi)\hat{k}_x(2p_y-\hat{k}_y)},\notag 
\end{gather}
\begin{align}
 v^{\rm HF}_{l_1,l_2,l_3,l_4}({\bf k})=&
 V({\bf k})
 f_{l_1,l_2}^0(-{\bf k})f_{l_3,l_4}^0({\bf k})-\int\frac{d^2k'}{(2\pi)^2}V({\bf k}')f_{l_1,l_4}^0(-{\bf k}')
 f_{l_3,l_2}^0({\bf k}')e^{-(i/2\pi)(k'_xk_y-k'_yk_x)}. 
\end{align}
and $f^0_{l_1,l_2}({\bf k})$ is given in Appendix \ref{app:LL_matrix}. 
In the definition of the HF potential 
$v^{\rm HF}_{l_1,l_2,l_3,l_4}({\bf k})$, 
the first term and the
second term in the right hand side are 
the Hartree term and the Fock term, 
respectively. 
The LL projected Hamiltonian Eq. (\ref{eq:LL_projected_Hamiltonian}) 
is obtained by projecting $\mathcal{H}_{\rm HF}$ into the $l$th LL. 

In Sec. \ref{sec:energy}, 
the deviations of the magnetic field and the two-point function 
are taken into account in the long
wavelength limit in order to consider the current effect on the HF
states. The deviation of the two-point function is caused by 
the deviation of $\langle \bar{\rho}_{l_1,l_2}(-\tilde{\bf k}) \rangle$. 
We only consider the deviation at the partially filled LL $l_0$ 
since it would give the largest contribution in our calculation 
and denote it as $\delta\bar\rho_{l_0}(-\tilde{\bf k})$.  
Then, the deviation of $\mathcal{H}_{\rm HF}$ is given by 
\begin{equation}
 \delta\mathcal{H}_{\rm HF}=\sum_{l_3,l_4}
 \int\frac{d^2k}{(2\pi)^2}v_{l_0,l_0,l_3,l_4}^{\rm HF}(\tilde{\bf k})
 \delta\bar{\rho}_{l_0}(-\tilde{\bf k})\bar{\rho}_{l_3,l_4}(\tilde{\bf k}). 
\end{equation} 
In the long wavelength limit, 
$\delta\bar\rho_{l_0}(-\tilde{\bf k})$ is relevant only for the small
momentum. 
When we expand $v_{l_1,l_2,l_3,l_4}^{\rm HF}(\tilde{\bf k})$ with respect to
${\bf k}$, 
the largest contribution comes from the lowest order term  
with respect to ${\bf k}$. 
For each set of the LLs $(l_3,l_4)$, 
the Hartree term of the HF potential gives the lower order term with respect
to ${\bf k}$ since the Hartree term has $V({\bf k})$ which is
$\mathcal{O}(1/k)$. Hence, the Hartree term gives the main contribution 
and the Fock term is negligible in the long wavelength limit. 

\section{Self-consistent equation for the ACDW state}
\label{app:self_eq}
The ACDW state at $\nu^\ast=1/2$ is constructed using the mean value of 
the projected density operator given
by Eq. (\ref{eq:projected_density_of_ACDW}). 
From Eq. (\ref{eq:projected_density_of_ACDW}), 
the two-point function of the operator 
$b_l({\bf p})$ is obtained by 
\begin{align}
 \langle  b_l^\dag({\bf p}) b_{l'}({\bf p'})\rangle_{\rm ACDW}
 &=
 \delta_{l,l'}\sum_{\bf N}e^{-i\phi({\bf p},{\bf N})}\big\{
 F_0({\bf p})(2\pi)^2\delta^2({\bf p}-{\bf p'}-2\pi {\bf N})\nonumber  \\
 &+F_1({\bf p})(2\pi)^2
 \delta(p_x-p'_x-2\pi N_x)\delta(p_y-p'_y-\pi(2N_y+1))\big\}, 
 \label{eq:self_eq_for_b}
\end{align}
where 
\begin{align}
 F_0({\bf p})=\sum_{\bf N}&
 \Delta_l(\frac{2\pi N_x}{r_s}, 2\pi N_y r_s)e^{-ip_x N_y+ip_y N_x-i\pi(N_x+N_y+N_xN_y)},\nonumber  \\
 F_1({\bf p})=\sum_{\bf N}&\Delta_l(\frac{2\pi N_x}{r_s},
 \pi(2N_y+1)r_s)\times e^{-ip_x N_y+ip_y N_x-i\pi(N_x+N_y+N_x(N_y+1/2))}. 
\end{align} 
The HF Hamiltonian of the ACDW state is given by
Eq. (\ref{eq:HF-ACDW}) and 
the ground state is 
the state in which the lower energy band 
at the uppermost partially-filled LL $l$ is fully occupied. 
The ground state is expressed in terms of the field operator 
$c_-({\bf p})$ by 
$|\Omega \rangle=N_c\prod_{{\bf p}\in {\rm RBZ}} c_-^\dag({\bf p})|0\rangle$, 
where $N_c$ is a normalization constant and 
$|0\rangle$ is a vacuum state 
in which the $(l-1)$th and lower Landau levels are fully occupied. 
The self-consisntent equation for $\Delta_l({\bf Q}_N)$ 
is obtained by calculating the 
the left hand side of Eq. (\ref{eq:self_eq_for_b}) 
for this ground state. 

Assuming the $x$- and $y$-inversion symmetries for the density, 
the order parameters become real and have the property 
$\Delta_l(Q_x,Q_y)=\Delta_l(Q_x,-Q_y)=\Delta_l(-Q_x,Q_y)$. 
The self-consistent solution is available 
only when the two energy bands are symmetric with
respect to the energy gap i.e., 
$\epsilon_+({\bf p})=-\epsilon_-({\bf p})\equiv \epsilon({\bf p})$, as expected from 
the particle-hole symmetry of the original Hamiltonian. 
This gives ${\rm Tr}_{2\times 2}D_l({\bf p})=0$, where 
${\rm Tr}_{2\times 2}$ denotes the trace with respect to the 
$2\times 2$ matrix indices, and 
$A(p_x, p_y+\pi)=-A({\bf p})$. 
In this case, $U({\bf p})$ and $\epsilon({\bf p})$ are given by 
\begin{gather}
 U({\bf p})=
 \left(
  \begin{array}{cc}
   \frac{B({\bf p})}{N_+({\bf p})} &\frac{B({\bf p})}{N_- ({\bf p})} \\
   \frac{\epsilon_+ ({\bf p})-A({\bf p})}{N_+ ({\bf p})} & 
    \frac{\epsilon_- ({\bf p})-A({\bf p})}{N_- ({\bf p})}
  \end{array}
 \right), \notag \\
 \epsilon({\bf p})=\sqrt{(A({\bf p}))^2+|B({\bf p})|^2}, 
\end{gather}
where 
$N_\pm({\bf p})= 2\epsilon_\pm({\bf p})(\epsilon_\pm({\bf p})-A({\bf p}))$.

\section{Landau level matrix elements}
\label{app:LL_matrix}
The matrix elements $\langle l_1 |e^{i{\bf q}\cdot {\bm \xi}}|l_2
\rangle$ are given as follows: 
\begin{widetext}
\begin{equation}
 \langle l_1 |e^{i {\bf q}\cdot{\bm \xi}}| l_2
  \rangle\\
 =
  \left\{
   \begin{array}{ll}
    \sqrt{\frac{l_1!}{l_2!}}
     \left(
      \frac{q_x+iq_y}{\sqrt{4\pi}}
     \right)^{l_2-l_1}e^{-\frac{q^2}{8\pi}}L_{l_1}^{l_2-l_1}
     \left(
      \frac{q^2}{4\pi}
     \right)
     & \hbox{for $l_2>l_1$}
     \\
    \sqrt{\frac{l_2!}{l_1!}}
     \left(
      \frac{q_x-iq_y}{\sqrt{4\pi}}
     \right)^{l_1-l_2}e^{-\frac{q^2}{8\pi}}L_{l_2}^{l_1-l_2}
     \left(
      \frac{q^2}{4\pi}
      \right)
     & \hbox{for $l_1>l_2$}\\
    e^{-\frac{q^2}{8\pi}}L_{l_1}
     \left(
      \frac{q^2}{4\pi}
       \right)
     & \hbox{for $l_2=l_1$},
   \end{array}
       \right.. 
  \label{eq:def_of_LL}
\end{equation}
From Eq.(\ref{eq:def_of_LL}), 
we obtain $f^\mu_{l_1,l_2}({\bf q})$ defined by Eq.(\ref{eq:def_of_f}) as 
$f^0_{l_1, l_2}({\bf q})=
\langle l_1|e^{i{\bf q}\cdot {\bm \xi}}|l_2\rangle$, 
$f^x_{l_1,l_2}({\bf q})=i\omega_c\partial_{q_y}
\langle l_1|e^{i{\bf q}\cdot {\bm \xi}}|l_2\rangle$ 
and 
$f^y_{l_1,l_2}({\bf q})=-i\omega_c\partial_{q_x}
\langle l_1|e^{i{\bf q}\cdot {\bm \xi}}|l_2\rangle$. 
Note that $\{f^\mu_{l_1,l_2}(-k)\}^\ast=f^\mu_{l_2,l_1}(k)$ holds 
following from its definition. 
The values of $f^\mu_{l_1,l_2}(0)$ and its derivatives are given as 
\begin{align}
&f_{l_1,l_2}^{0}(0)=\delta_{l_1,l_2},\nonumber \\
&f_{l_1,l_2}^{x}(0)=\left.
i\omega_c \frac{\partial f_{l_1,l_2}^{0}(q)}{\partial q_y}\right|_{q=0}
=
\left\{
\begin{array}{ll}
-\omega_c \sqrt{\frac{l_1+1}{4\pi}}
\delta_{l_2,l_1+1}& \hbox{for $l_2>l_1$}\\
\omega_c\sqrt{\frac{l_1}{4\pi}}
\delta_{l_1,l_2+1}& \hbox{for $l_1>l_2$}\\
0 & \hbox{for $l_1=l_2$}
\end{array}
\right. ,\\
&f_{l_1,l_2}^{y}(0)=
\left.
-i\omega_c \frac{\partial f_{l_1,l_2}^{0}(q)}{\partial q_x}\right|_{q=0}
=
\left\{
\begin{array}{ll}
-i\omega_c\sqrt{\frac{l_1+1}{4\pi}}
\delta_{l_2,l_1+1}& \hbox{for $l_2>l_1$}\\
-i\omega_c\sqrt{\frac{l_1}{4\pi}}
\delta_{l_1,l_2+1}& \hbox{for $l_1>l_2$}\\
0 & \hbox{for $l_1=l_2$}
\end{array}
\right., 
\end{align}
\begin{align}
 &\left.
 \frac{\partial f_{{l_1,l_2}}^{x}(q)}{\partial q_y}
 \right|_{q=0}
 =
 \left\{
 \begin{array}{ll}
  -i\frac{\omega_c}{4\pi}
   \sqrt{(l_1+1)(l_1+2)}\delta_{l_2,l_1+2}
   &  \hbox{for $l_2>l_1$}\\
  -i\frac{\omega_c}{4\pi}
   \sqrt{l_1(l_1-1)}\delta_{l_1,l_2+2}
   &  \hbox{for $l_1>l_2$}\\
  -i\frac{\omega_c}{4\pi}(l_1+\frac{1}{2}) & \hbox{for $l_1=l_2$}\\
 \end{array}
 \right. ,\\
 &\left. 
 \frac{\partial f_{{l_1,l_2}}^{y}(q)}{\partial q_x}\right|_{q=0}
 =
 \left\{
 \begin{array}{ll}
  -i\frac{\omega_c}{4\pi}
   \sqrt{(l_1+1)(l_1+2)}\delta_{l_2,l_1+2}
   &  \hbox{for $l_2>l_1$}\\
  -i\frac{\omega_c}{4\pi}
   \sqrt{l_1(l_1-1)}\delta_{l_1,l_2+2}
   &  \hbox{for $l_1>l_2$}\\
  i\frac{\omega_c}{4\pi}(l_1+\frac{1}{2})
   & \hbox{for $l_1=l_2$}
 \end{array}
 \right..
\end{align}
\end{widetext}

\section{Calculation of $K^{00}$ for the ACDW state}
\label{app:K00}
When $p_0=p_y=0$ and $p_x\to 0$, the response function $K^{00}_0(p_x)$
is Taylor expanded with respect to $p_x$ as 
\begin{equation}
 K^{00}_0(p_x)=K^{00}_0(0)+p_x\partial_{p_x}K^{00}_0(0)+
  \frac{p^2_x}{2}\partial^2_{p_x}K^{00}_{0}(0)+\dots
\end{equation} 
The first and second terms become zero. 
The third term includes the corrections from the inter-LL term and the
intra-LL term. The inter-LL term gives the same expression for
$K^{00}_0$ as that in the striped Hall state. 
The intra-LL term gives the extra correction $\Delta K^{00}_0(p_x)$
given by 
\begin{equation}
\Delta K^{00}_0(p_x)=\kappa \times p^2_x, 
\end{equation}
where $\kappa$ is given by 
\begin{widetext}
\begin{equation}
 \kappa=\frac{e^2}{2}
  \int_{\rm RBZ}\frac{d^2p'}{(2\pi)^2}\frac{1}{2\epsilon({\bf p'})}
  \partial^2_{p_x}
  \left. \left[
   \frac{A(\hat{\bf p}+{\bf p}')A({\bf p}')+
   {\rm Re}(B(\hat{\bf p}+{\bf p}')B^\ast({\bf p}')e^{-(i/2)\hat{p}_x})}
   {\epsilon(\hat{\bf p}+{\bf p}')\epsilon({\bf p}')}
  \right]
  \right|_{p_x=0}.
\end{equation}
\end{widetext}
$\beta$ in Eq.(\ref{eq:K_for_ACDW}) is defined by 
$\beta\equiv - \kappa \times \nu \omega_c/\sigma_{xy}^{(\nu)} \sqrt{B}$. 
The finite $\kappa$ is the result of the band formation at 
the partiall filled LL, 
while the band structure is generated by 
the density modulation of the ACDW state in both directions. 
Hence, it may be considered that 
the finite $\kappa$ reflects 
the remaining density modulation effect of the ACDW state 
in the long wavelength limit.


\end{document}